\documentclass[fleqn,usenatbib]{mnras}
\pdfoutput=1
\usepackage{newtxtext,newtxmath}
\usepackage[T1]{fontenc}
\usepackage{ae,aecompl}

\usepackage{graphicx}	
\usepackage{epsfig}
\usepackage{float}
\floatstyle{ruled}
\restylefloat{table}

\usepackage{amsmath}	
\usepackage{amssymb}	
\usepackage[titletoc]{appendix}
\usepackage{booktabs}
\usepackage{rotating} 

\title[A dynamical mass proxy for galaxies across the Hubble sequence.]{Kinematic scaling relations of CALIFA galaxies: A dynamical mass proxy for galaxies across the Hubble sequence.}

\author[E. Aquino-Ort\'iz et al.]{
E. Aquino-Ort\'iz,$^{1}$\thanks{E-mail: eaquino@astro.unam.mx}
O. Valenzuela$,^{1}$
S. F. S\'anchez,$^{1}$
H. Hern\'andez-Toledo,$^{1}$ 
\newauthor 
V. \'Avila-Reese,$^{1}$
G. van de Ven,$^{3,4}$
A. Rodr\'iguez-Puebla,$^{1}$
L. Zhu,$^{3}$
\newauthor 
B. Mancillas,$^{2}$
M. Cano-D\'iaz,$^{1}$
R. Garc\'ia-Benito.$^{5}$
\\
$^{1}$Instituto de Astronom\'ia, Universidad Nacional Aut\'onoma de M\'exico, A.P. 70-264, 04510 CDMX, Mexico.\\
$^{2}$LERMA, CNRS UMR 8112, Observatoire de Paris, 61 Avenue de l'Observatoire, F-75014 Paris, France.\\
$^{3}$Max Planck Institute for Astronomy, Konigstuhl 17, 69117 Heidelberg, Germany.\\
$^{4}$European Southern Observatory, Karl-Schwarzschild-Str. 2, 85748 Garching b. Munchen, Germany.\\
$^{5}$Instituto de Astrof\'isica de Andaluc\'ia (CSIC), P.O. Box 3004, 18080 Granada, Spain.\\
}

\date{Accepted 2018 June 6. Received 2018 May 26; in original form 2018 April 25}

\pubyear{2018}

\defcitealias{Cortese+2014}{C14}
\defcitealias{Falcon-Barroso+2017}{FB17}

\begin{document}
\label{firstpage}
\pagerange{\pageref{firstpage}--\pageref{lastpage}}
\maketitle

\begin{abstract}
We used ionized gas and stellar kinematics for 667 spatially resolved galaxies publicly available from the Calar Alto Legacy Integral Field Area survey (CALIFA) 3rd Data Release with the aim of studying kinematic scaling relations as the Tully $\&$ Fisher (TF) relation using rotation velocity, $V_{rot}$, the Faber $\&$ Jackson (FJ) relation using velocity dispersion, $\sigma$, and also a combination of $V_{rot}$ and $\sigma$ through the $S_{K}$ parameter defined as $S_{K}^2 = KV_{rot}^2 + \sigma^2$ with constant $K$. Late-type and early-type galaxies reproduce the TF and FJ relations. Some early-type galaxies also follow the TF relation and some late-type galaxies the FJ relation, but always with larger scatter. On the contrary, when we use the $S_{K}$ parameter, all galaxies, regardless of the morphological type, lie on the same scaling relation, showing a tight correlation with the total stellar mass, $M_\star$. Indeed, we find that the scatter in this relation is smaller or equal to that of the TF and FJ relations. We explore different values of the $K$ parameter without significant differences (slope and scatter) in our final results  with respect the case $K=0.5$ besides than a small change in the zero point. We calibrate the kinematic $S_{K}^2$ dynamical mass proxy in order to make it consistent with sophisticated published dynamical models  within $0.15\ dex$. We show that the $S_{K}$  proxy is able to reproduce the relation between the dynamical mass and the stellar mass in the inner regions of galaxies. Our result may be useful in order to produce fast estimations of the central dynamical mass in galaxies and to study correlations in large galaxy surveys.
\end{abstract}




\section{Introduction.}

\begin{figure*}
\centering
\epsfig{file=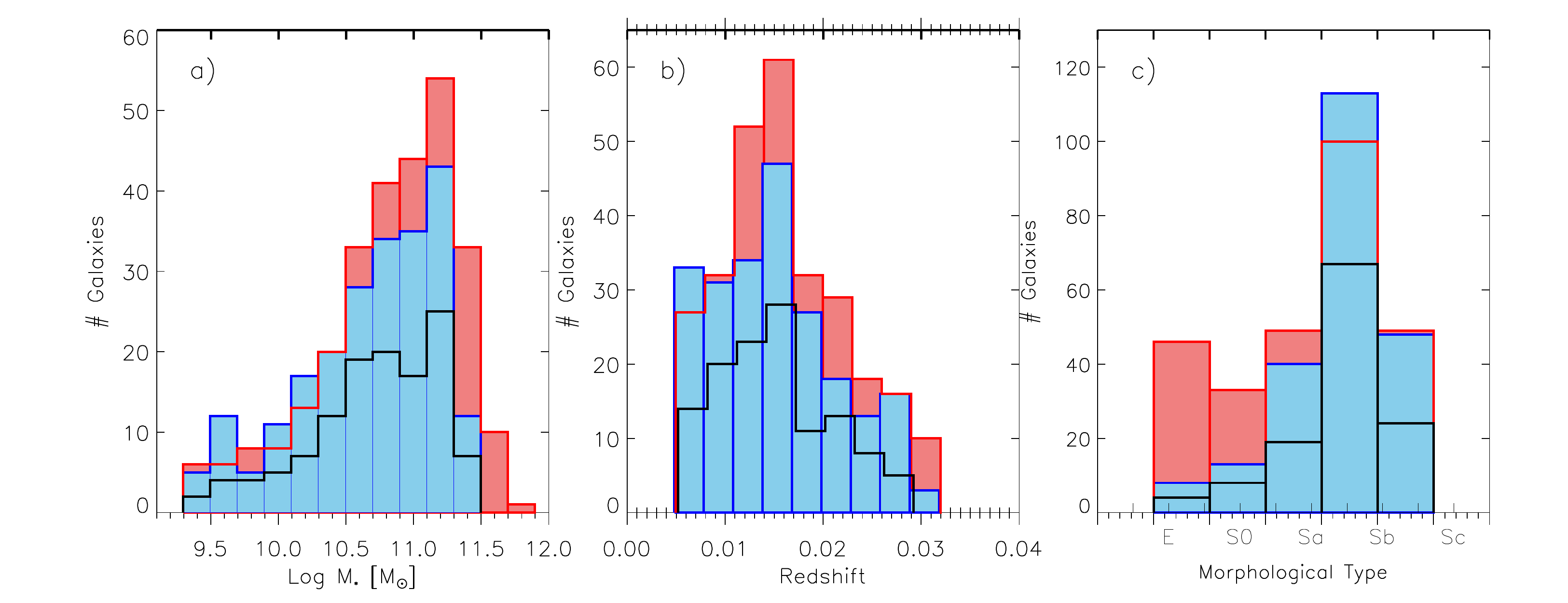,width=18cm, height=6cm}
\caption{Galaxy Sample distributions in (a) total stellar masses, (b) redshift and (c) morphological type. Blue and red histograms indicate galaxies with ionized gas and stellar kinematics, respectively, whereas the unfilled black histogram indicate galaxies with both, ionized gas and stellar kinematics.}
\label{fig:histograms}
\end{figure*}

Galaxy scaling relations describe trends that are observed between different properties of galaxies. They are assumed to be the consequence of their formation and evolution. Probably the kinematic scaling relation most widely studied for spiral galaxies is the Tully-Fisher relation (hereafter TF); a correlation between luminosity and rotational velocity, firstly reported by \citet{Tully&Fisher1977}. It was originally established as a tool to measure distances to spiral galaxies \citep[e.g.,][]{Giovanelli+1997}. It has been suggested that the slope, zero-point and tightness may have a cosmological origin helping us to understand the formation and evolution of galaxies \citep[e.g][]{Cole+1994,Eisenstein+1996,Mo+1998,Avila-Reese+1998,Courteau1999,Firmani2000,Navarro+2000}. In the local universe the TF relation is very tight \citep[e.g.][]{Verheijen2001, Bekerait+2016, Ponomareva+2017}, locating galaxies with rising rotation curves on the low-velocity end and galaxies with declining rotation curve on the high-velocity end \citep[e.g.,][]{persic96}. The luminosity-based TF is more directly accessible, however, the amount of light measured from the stellar population is a function of passband, and therefore different TF relations emerge when observing galaxies at different wavelengths. A physically more fundamental approach instead of luminosity is based on stellar mass, $M_\star$. The resulting TF relation is well approximated by a single power law with small scatter at least for disk galaxies more massive than $\sim 10^{9.5} M_{\odot}$ \citep[e.g.][]{McGaugh+2000, Bell&deJong2001, Avila-Reese+2008}. 
A similar correlation between the luminosity (or the stellar mass) of elliptical galaxies and the velocity dispersion in their central regions was established by \citet{Faber&Jackson1976} (hereafter FJ). The shape and scatter of the FJ relation has been less frequently studied because its large residuals show a significant correlation with galaxy size, i.e., a third parameter within the so called fundamental plane \citep[e.g.][]{Djorgovski&Davis1987, Dressler+1987, Cappellari+2013, Desmond&Wechsler2017}.

It is presumed that galaxy internal kinematics as tracer of the gravitational potential provide the dynamical mass. If spiral and elliptical galaxies were completely dominated by rotation velocity and velocity dispersion, respectively, the TF and FJ relations would provide insights into the connection between galaxies and their dark matter content. However, structural properties, environmental effects or internal physical processes perturb the kinematics of late-type galaxies producing non-circular motions that under/overestimates the circular velocity \citep[e.g.][]{Valenzuela+2007, Randriamampandry+2015, Holmes+2015}. On the other hand, elliptical galaxies, although dominated by velocity dispersion, often present some degree of rotation \citep[e.g][]{Lorenzi+2006,Emsellem+2007,Emsellem+2011,Cappellari+2011,Rong2018}.
Non-circular motions on disk galaxies and rotation on ellipticals may contribute to miss a fraction of the gravitational potential, modifying the scaling relations and precluding them from being directly comparable to theoretical predictions.

\citet{Weiner+2006} introduced a new kinematic parameter involving a combination of rotation velocity and velocity dispersion in order to study high redshift galaxies, where in some cases random motions were not negligible. \citet{Weiner+2006} showed that such parameter provides a better proxy to the integrated line-width of galaxies emission lines than rotation velocity or velocity dispersion alone, regardless of the galaxy morphology. The parameter is defined as:
\begin{equation}
S_{K}^{2} = KV_{rot}^{2} + \sigma^{2},
\end{equation}
where $V_{rot}$ is the rotation velocity, $\sigma$ is the velocity dispersion and $K$ a constant that could be extremely complicated function of the formation history, dynamic state and environment of galaxies.
\citet{Kassin+2007} found that by adopting a value of $K=0.5$, the $S_{0.5}$ parameter presents a tight correlation with the stellar mass for a sample of galaxies at redshift $z \leq 1.2$ extracted from the All Wavelength Extended Groth Strip International Survey (AEGIS) and the Deep Extragalactic Evolutionary Probe 2 (DEEP2). This correlation seems to be independent of the morphological type. Other analyses, focused on the evolution of the TF relation at high redshift ($z\sim 2$), explored the $M_\star-S_{0.5}$ relation confirming that turbulent motions might play an important dynamical role \citep[e.g.][]{Cresci+2009, Gnerucci+2011, Vergani+2012, Price+2016, Christensen2017}. \citet[][]{Zaritsky+2008} provided a possible explanation of the $M_\star-S_{0.5}$ relation as a virial one, includying all galaxy evolution, geometrical and dynamical complications into the K coefficient.

\citet[][hereafter \citetalias{Cortese+2014}]{Cortese+2014} performed the only systematic study of this relation at low redshift ($z \leq 0.095$). They used the stellar and ionized gas kinematics integrated within one effective radius, $r_{e}$, for galaxies observed with the Sydney-AAO Multi-object Integral Field survey \citep[SAMI;][]{sami}. \citetalias{Cortese+2014} confirmed that all galaxies, regardless of the morphological type, lie on the same kinematic scaling relation $M_\star-S_{0.5}$ with a significant improvement compared with the TF and FJ relations. Although the result is encouraging, the spatial covering of the observations ($1r_{e}$) and the coarse spatial resolution of the data may contribute to the uncertainties in a similar way as they do it in HI line-width TF estimations \citep[e.g.][]{Ponomareva+2017}. Therefore, it is needed to repeat this analysis using data with better spatial resolutions and coverage.

The aim of this paper is to explore and calibrate the TF, FJ and $S_{K}$ scaling relations in the local universe for galaxies from the Calar Alto Legacy Integral Field Area survey \citep[CALIFA;][]{Sanchez+2012}. These data present a larger spatial coverage and better physical resolution \citep{DR3}.\footnote{Both surveys present a similar projected PSF FWHM of $\sim$2.5$\arcsec$. However CALIFA sample galaxies observed in a considerable lower redshift and narrower redshift range.} In a recent study, Gilhuly et al. (submitted), presented an exploratory study of these relations for a limited sample of galaxies. They perform a systematic and detailed analysis of the limitations of the kinematics parameters, and in particular the velocity dispersion in the CALIFA dataset. Finally, they present the largest and more precise estimation of individual velocity dispersions based on CALIFA observations, as a public accessible table. The current study would explore  the nature of these scaling relations.

The structure of this article is as follows. In Section \ref{sec:sample} we briefly
describe the CALIFA sample, including a summary of the delivered datasets. Details of the analysis perform over the data are presented in Section \ref{sec:spec}. In Section \ref{sec:int_kin} we estimate the kinematics parameters within $1r_{e}$, following the same methodology as \citetalias{Cortese+2014}. In Section \ref{sec:res_kin} we perform a detailed modeling of the 2D spatially resolved velocity maps for a subsample of good quality datasets. With this modeling we estimate the possible effects of aperture and non circular motions in disk galaxies and obtain a more precise derivation of the maximum rotational velocity, $V_{max}$. In Section \ref{sec:results}, we present the main results of this study. In Section \ref{sec:Disc} we discuss the results and their physical implications and finally we summarize the main conclusions in Section \ref{sec:Concl}.

\section{Data sample.}
\label{sec:sample}

We make use of the data provided by the Calar Alto Legacy Integral Field Area (CALIFA) survey \citep{sanchez12a}, that has delivered publicly available integral field spectroscopy data for 667 galaxies \citep{DR3}, although the current samples comprises more than 700 galaxies \citep{sanchez17}. Details of the observational strategy and data reduction are explained in these two articles. All galaxies were observed using PMAS \citep{roth05} in the PPaK configuration
\citep{kelz06}, covering an hexagonal field of view (FoV) of 74$\arcsec$$\times$64$\arcsec$, which is sufficient to map the full optical extension of most of the galaxies up to two to three effective radii. This is possible due to the diameter selection of the CALIFA sample \citep{walcher+2014}. The final observed sample comprises galaxies of any morphological type (See Figure \ref{fig:histograms}). It covers, with a good sampling, the color-magnitude diagram and the stellar mass distributions of the Local Universe in a representative and statistically significant way for galaxies more massive than 10$^{9.5}$ M$_\odot$ \citep[e.g.][]{walcher+2014,DR3}

The observing strategy guarantees a complete coverage of the FoV, with a final spatial resolution of FWHM$\sim$2.5$\arcsec$, corresponding to $\sim$1 kpc at the average redshift of the survey \citep[e.g][]{rgb15,DR3}. CALIFA observed the galaxies in two configurations: (i) the V500 setup, a low resolution mode that covers a wavelength range between 3745-7500 \AA, with a resolution of $\lambda/\Delta\lambda\sim$850 (FWHM$\sim$6\AA),
 and (ii) the V1200 setup, an intermediate resolution mode, that covers the wavelength range between 3700-4800\AA, with a resolution of $\lambda/\Delta\lambda\sim$1650 (FWHM$\sim$2.7\AA). The delivered dataset was reduced using version 2.2 of the CALIFA pipeline, whose modifications with respect to the previous ones \citep{sanchez12a,husemann13,rgb15} are described in \citet{DR3}. The final dataproduct of the reduction is a datacube comprising the spatial information in the {\it x} and {\it y} axis, and the spectral one in the {\it z} one. For further details of the adopted dataformat and the quality of the data consult \citet{DR3}.

\section{Analysis.}
\label{sec:ana}

We describe here the analysis performed to estimate the stellar mass distribution and the kinematics parameters for the different galaxies included in the current dataset.

\subsection{Spectroscopic Analysis.}
\label{sec:spec}

In this paper we use the data-products (ionized gas kinematic maps) derived for the CALIFA V500 setup dataset by Pipe3D pipeline \citep{Sanchez2016} based on the Fit3D fitting tool \citep{Sanchez2015}, together with the stellar line-of-sight velocity and intrinsic dispersion maps for the V1200 setup performed using pPXF by \citet{Falcon-Barroso+2017}.

Pipe3D models the stellar continuum adopting a multi Single Stellar Population (SSP) template library, taking into account the velocity, dispersion and dust attenuation of the stellar populations. Then, it estimates the main properties of the nebular emission lines. The current implementation of Pipe3D adopted the GSD156 \citep{Cid_Fernandes+2013} template library for the analysis of the stellar population properties. This library comprises 156 templates covering 39 stellar ages (from 1Myr to 13Gyr), and 4 metallicities ($Z$/$Z_{\odot}$ = 0.2, 0.4, 1, and 1.5). A spatial binning for the stellar population analysis was applied to reach an homogeneous S/N of 50 across the field of view. The stellar population fitting was applied to the co-added spectra within each spatial bin. Finally, following the procedures described in \citet{Cid_Fernandes+2013} and \citet{Sanches2016a}, was estimated the stellar-population model for each spaxel by re-scaling the best-fit model within each spatial bin to the continuum flux intensity in the corresponding spaxel. The stellar-population model spectra are then subtracted from the original datacube to create a gas-pure cube comprising only the ionized gas emission lines. For this pure-gas cube, the stronger emission lines were then fitted spaxel by spaxel using single Gaussian models for each emission line in each individual spectrum to derive the corresponding flux intensity and line-of-sight kinematics.
In addition, the spatial distribution of the stellar mass densities and the integrated stellar masses at different apertures are recovered from the Pipe3D analysis by taking into account the decomposition in SSPs, the Mass-to-Light ratio of each of them, and the integrated light at each spaxel within the FoV. For this derivation was assumed the Salpeter Initial Mass Function  \citep{Salpeter:1955p3438}.
More details of the fitting procedure, adopted dust attenuation curve, and uncertainties of the process are given in \citet{Sanchez2015,Sanchez2016}.

\citet{Falcon-Barroso+2017} performed a detailed analysis to extract the stellar kinematics for the intermediate resolution CALIFA data (V1200 setup). The data-cubes were spatially binned with the Voronoi 2D binning method of \citet{cappellari03} to achieve an approximately constant signal-to-noise (S/N) of 20 per spaxel taking in account the correlation in the error spectrum of nearby spaxels \citep[see][for details]{husemann13}. This S/N value conserve a good spatial resolution while still being able to reliably estimate the line-of-sight velocity distribution. The stellar kinematics was estimated using the pPXF code of \citet{Cappellari&Emsellem2004}. The stellar templates were taken from the Indo-US spectral library \citep{Valdes+2004} with $\sim 330$ selected stars. The stellar rotation velocity, the velocity dispersion and corresponding error were estimated by $\chi^{2}$ minimization in pixel space as the bi-weight mean and standard deviations of a set of 100 Monte Carlo realizations of the fitting.

\subsection{Integrated kinematics.}
\label{sec:int_kin}

The original dataset comprises 734 galaxies for the V500 dataset observed within the frame-work of the CALIFA survey \citet{sanchez17}, and the 300 galaxies for the V1200 dataset described by \citet{Falcon-Barroso+2017}. From this dataset we perform a selection of the optimal data for the proposed analysis following the methodology described by \citetalias{Cortese+2014}: First, spaxels are discarded if the error in velocities is greater than $20\ km/s$ and $50\ km/s$ for gaseous and stellar kinematics, respectively. This conventional cut correspond to one third of the spectral FWHM ($\sim 6\ $\AA, i.e.,\ $\sim 150\ km/s$) of V500 CALIFA data. 
Second, we selected only those galaxies for which at least 80\%\ of the spaxels within an ellipse of semi-major axis equal to $1r_e$ fulfill this quality criteria.
This criteria guarantee that we are tracing well the kinematics parameters up to $1r_{e}$. Finally, galaxies under merging and clear traces of interactions are discarded based on morphological distortions and the abundance of galaxy neighbors with a comparable size. Following this procedure our final sample comprise 223 galaxies with ionized gas kinematics (V500 setup), 278 with stellar kinematics (V1200 setup), and 123 with both of them.

\subsubsection{Velocity dispersion: $\sigma$.}

Stellar velocity dispersions were estimated as the linear average of the velocity dispersion of all spaxels within the ellipse mentioned in the previous section using the velocity dispersion maps from the V1200 dataset without correction for inclination. Following \citetalias{Cortese+2014} we use linear instead luminosity-weighted averages to be consistent with our velocity width measurements which are not luminosity-weighted. Ionized gas velocity dispersions were estimated fitting the integrated spectrum within a diameter of $5"$ with Pipe3D for the V500 dataset using the template library described above.
Regarding the determination of the stellar and gaseous velocity dispersions (up to $1r_{e}$), which dominate in early-type galaxies, we rely on the detailed kinematic analysis presented in \citet{Zhu+2017,Zhu2+2018}.

\subsubsection{Rotation velocity: $V_{rot}$.}

Once more, we followed \citetalias{Cortese+2014}, to derive the stellar (V1200 dataset) and gaseous (V500 dataset) rotation velocities. They adopted the same classical procedure developed to analyze the integrated HI emission profiles in galaxies, i.e., through the width parameter, $W$ \citep[e.g.][]{Mathewson+1992,Vogt+2004}. 
First, a histogram is derived of the velocities estimated for all the good spaxels within the $r_{e}$. Then, it is calculated the difference between the 10th and 90th percentile points of this velocity histogram, defined as the width: $W=V_{90} - V_{10}$ \citep[e.g.][]{Catinella+2005}. Finally, the rotation velocity is defined as:

\begin{equation}
V_{rot} = \frac{W}{2(1+z)sin(i)},
\label{Vrot_integrated}
\end{equation}
where $z$ is the redshift and $i$ is the galaxy inclination determined from the observed ellipticity $\varepsilon$ as:

\begin{equation}
cos(i) = \sqrt[]{\frac{(1-\varepsilon)^2 - q_{0}^2}{1 - q_{0}^2}},
\end{equation}
with $q_{0}$ being the intrinsic axial ratio of edge-on galaxies. Following \citet{Catinella+2012} and \citetalias{Cortese+2014}, we adopted $q_{0}=0.2$ for all galaxies and set the inclination to $90^\circ$ edge-on if $\varepsilon\geq0.8$.

Integrated rotation velocity estimated by Eq. \ref{Vrot_integrated} is a good representation of the maximum rotation velocity, $V_{max}$, if the kinematics of the galaxy is axi-symmetric (i.e., without non-circular motions).  However, this is not always the case. Some galaxies show deviations from a pure rotational pattern due to warps, lopsidedness, arms, bars, outflows, inflows, nuclear activity, etc., \citep[e.g.][]{Bosma1978, Schoenmakers1997, Verheijen2001, Holmes+2015, Kalinova+2017, Sanchez-Menguiano+2017} producing non-circular motions and distorting the velocity profile (i.e., velocity histogram). In the next subsection, we try to quantify these effects in the derivation of $V_{max}$ by performing a more detailed analysis on a limited sample of galaxies and comparing the results.

\subsection{Spatially resolved kinematics: $V_{max}.$}
\label{sec:res_kin}

Kinematic maps of spiral galaxies are often treated as being consistent with a purely circular flow pattern. This means that the kinematics of a galactic disk at a certain galactocentric radius can be described by a single tilted ring model defined by three parameters: the rotation velocity and two parameters that describe the local disk orientation with respect to some reference system \citep[e.g.][]{Rogstad+1974}. Several routines exists to fit kinematic maps based in this method. The most extensively used is the ROTCUR routine \citep[e.g.][]{Begeman1989}, which fits a set of inclined rings to a velocity field.  However, as we mentioned above, the kinematics could be affected by the presence of non-circular motions and in some cases the tilted-ring model is an oversimplification.
A more precise kinematic analysis requires tools that consider non-circular motions.

\citet{Spekkens&Sellwood2007} and \citet{Sellwood&Sanchez2010} developed the VELFIT code specifically to characterize the non-circular motions in the kinematics of spiral galaxies expressed in a Fourier series. We used this code with some improvements (Aquino-Ort\'\i z in prep.) to derive the properties of the velocity maps. 
This fit provides an estimate of the rotation curve, the kinematic inclination and position angle of the galaxy, together with the amplitude of the non-circular motions as a function of radius. A bootstrap procedure is adopted to  estimate the uncertainties on the derived parameters.

\begin{figure*}
\centering
\epsfig{file=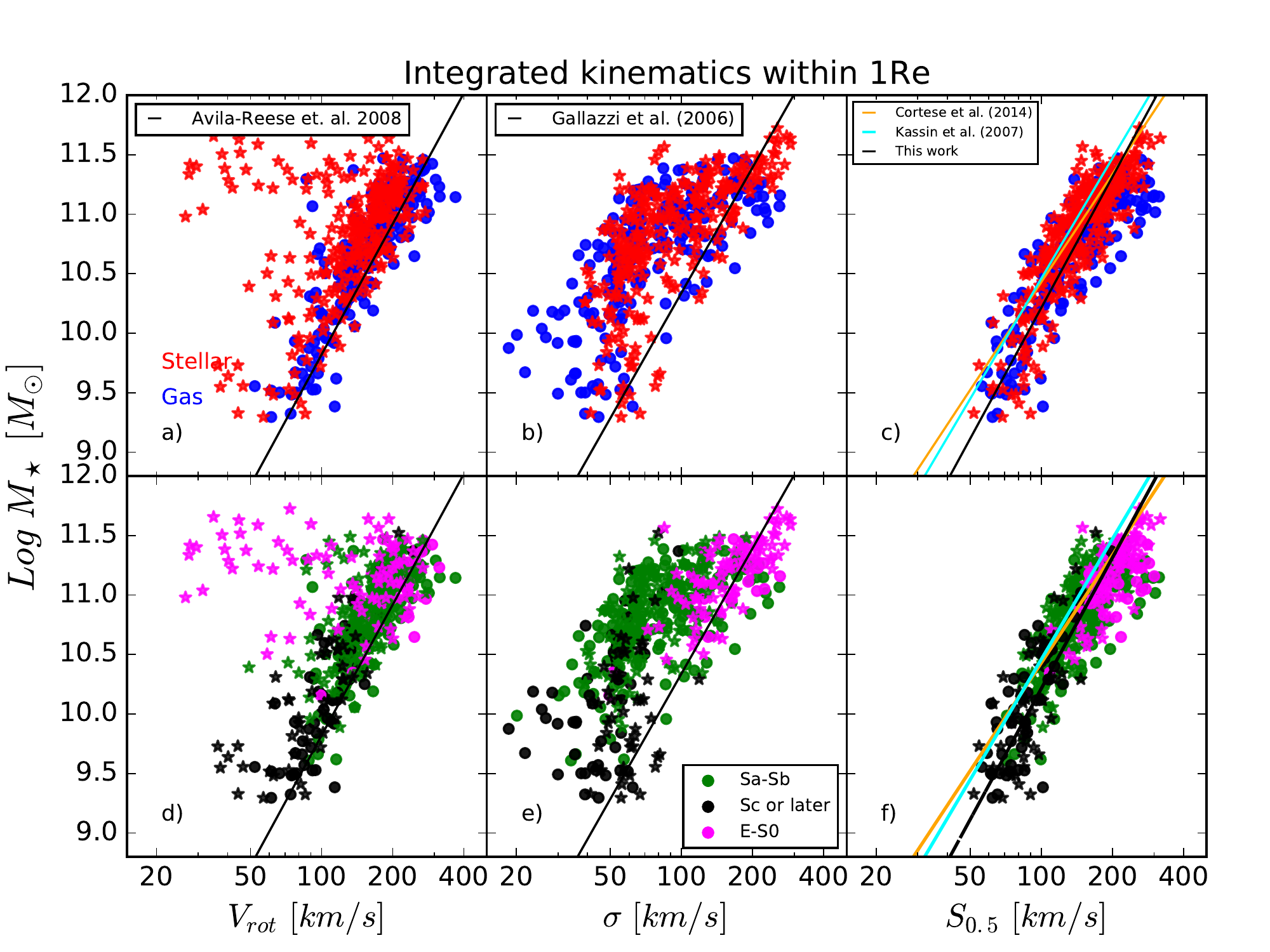,width=16cm, height=10cm}
\caption{Kinematic scaling relations with integrated kinematics. \textit{Left:} Tully \& Fisher (TF) relation with the black line representing the orthogonal best-fit TF relation from \protect\citet{Avila-Reese+2008}. \textit{Middle:} Faber \& Jackson (FJ) relation with the black line the best-fit FJ relation from \protect\citet{Gallazzi+2006}. \textit{Right:} The $M_\star-S_{0.5}$ relation, cyan and yellow lines indicate the best-fit $M_\star-S_{0.5}$ relation from \protect\citet{Kassin+2007} and \protect\citet{Cortese+2014}, respectively, whereas the black line represent our best-fit. \textit{Top panels:} red star and blue circles represent galaxies with stellar and ionized gas kinematics. \textit{Bottom panels:} galaxies with different morphological types; magenta indicate elliptical and lenticular galaxies, green are Sa and Sb galaxies and black symbols are Sc galaxies.}
\label{fig:scaling_relations}
\end{figure*}

The current procedure is not performed over the full dataset, since in many cases the kinematics present clear deviations due to external perturbations or is strongly affected by random motions. We discarded those cases whose kinematics appeared highly disturbed by the presence of large nearby companions or clear indications of being in a merging process. Therefore, we select a control sample, with good quality spatial resolved kinematics, comprising those isolated galaxies with low velocity dispersion, and inclinations between $30^\circ < i < 70^\circ $. This sample of galaxies, that are the best suited for modeling their velocity maps, comprises 42 galaxies for ionized gas kinematics (V500) and 92 galaxies for stellar kinematics (V1200). 

The estimated rotation curves for all these galaxies present a great diversity, in agreement with previous results \citep[e.g.][]{Kalinova+2017}. For a limited fraction of galaxies $(\sim 10\%)$ the spatial coverage was insufficient to measure the maximum velocity, $V_{max}$. In order to still estimate $V_{max}$ we follow \citet{Bekerait+2016} and parametrize the rotation curve using the formula proposed by \citet{Bertola+1991}:

\begin{equation}
v(r) = v_{0} + \frac{v_{c}r}{(r^2 + k^2)^{\frac{\gamma}{2}}},
\end{equation}
where: (i) $v_{0}$ is the systemic velocity of the galaxy; (ii) $v_{c}$ is a parameter governing the amplitude of the rotation curve; (iii) $k$ describes its sharpness and finally (iv) $\gamma$ allows modeling rising or falling curves, with $\gamma = 1$ for a flat rotation curve.

\begin{figure*}
\centering
\epsfig{file=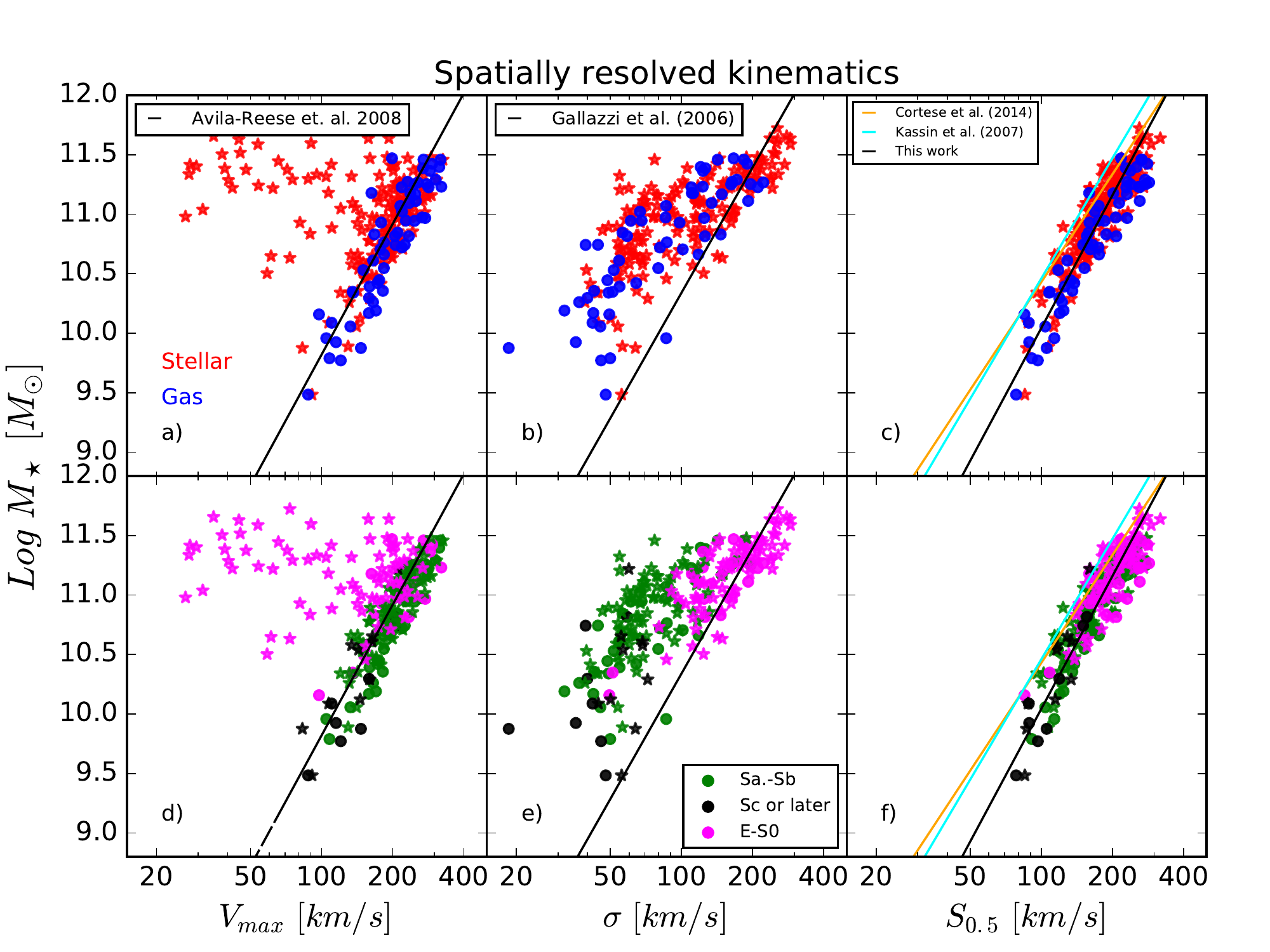,width=16cm, height=10cm}
\caption{Kinematic scaling relations for our control sample with spatially resolved kinematics. \textit{left)} Tully \& Fisher with the black line as the stellar mass Tully \& Fisher from \protect\citet{Avila-Reese+2008}. \textit{middle)} Faber \& Jackson with the black line as the \protect\citet{Gallazzi+2006}. \textit{right)} The $M_\star-S_{0.5}$ relation, cyan and yellow lines are the best-fit of \protect\citet{Kassin+2007} and \protect\citet{Cortese+2014}, respectively, whereas the black line represent our best-fit. \textit{Top panels:} red stars and blue circles represent galaxies with stellar and gas kinematics. \textit{Bottom panels:} the color-code represent different morphologies.}
\label{fig:SK_V1200}
\end{figure*}

\section{Results.}
\label{sec:results}

Figure \ref{fig:scaling_relations} shows the analyzed kinematic scaling relations using the total stellar mass (unless noted otherwise) and the integrated kinematics, segregated by stellar and ionized gas kinematics (upper panels) and by morphology (lower panels), respectively. In each panel is included some reference relations found by previous results, and the best-fit relations for the $M_\star-S_{0.5}$ distribution. Figure \ref{fig:SK_V1200} shows the same distributions for the resolved kinematics.

Table \ref{table} summarizes the results of an orthogonal linear fit along the horizontal axis, considering the total stellar mass on the vertical axis as the independent variable, using the routines presented by \citet{Akritas&Bershady1996}, for each of the kinematic scaling relations and dataset. It includes the zero-points and slopes, together with the scatter around the best-fit relations estimated from the 'error-in-variable' of the corresponding fit as the standard deviation of residuals. In addition is listed the reference results for the TF, FJ and $S_{0.5}$ scaling relations extracted from the literature shown in Figures \ref{fig:scaling_relations} and \ref{fig:SK_V1200}.

\subsection{TF relation.}
\label{TFR}

The TF relation including early type galaxies based on the integrated analysis are shown in the left panels of Figure \ref{fig:scaling_relations}. These relations show a large scatter, $0.084\ dex$ in $\log V_{rot}$ for ionized gas kinematics and $0.20\ dex$ for stellar kinematics. The value for ionized gas kinematics is in agreement with the one reported for the luminosity TF relation estimated by \citet{Bekerait+2016}, $\sim$ 0.09 dex, despite the fact that their study was based on a detailed analysis of the rotation velocity of a sub-sample of the CALIFA galaxies. In that study they analyzed their velocity within a radius containing 83$\%$ of all light, $V_{opt}$. On the other hand, our scatter for stellar kinematics is lower than the one reported by \citetalias{Cortese+2014} for SAMI ($\sim$ 0.25 dex). The difference with this later study is not surprising because the SAMI sample is dominated by Sc low-mass galaxies, where the rotation curves are still rising at $1r_{e}$, being far from $V_{max}$, whereas our sample is dominated by Sa and Sb galaxies (see lower panels of Figure \ref{fig:scaling_relations}).

Figure \ref{fig:SK_V1200}, left-panel, shows the TF relation also including early type galaxies based on the spatially resolved analysis (i.e, using $V_{max}$). The parameters of the best-fit relation to these data are listed in Table \ref{table}. When adopting this improved estimation of the velocity, the scatter decreases to $\sim$ 0.07 dex for the ionized gas kinematics, but it increases to $\sim$ 0.24 dex for the stellar one. This later value agrees with the one reported by \citetalias{Cortese+2014}. The scatter for our stellar kinematics TF relation increases due to that late-type galaxies move to higher velocities and also for the inclusion of early-type galaxies in the relation. Such galaxies are undetected in the gas component, being  the analyzed sample limited to mostly late-type galaxies.

As a reference we include in Figure \ref{fig:scaling_relations} and Figure \ref{fig:SK_V1200} the derivation of the stellar TF relation as presented in \citet{Avila-Reese+2008}\footnote{We have increased the stellar mass in \citet{Avila-Reese+2008} by $0.15\ dex$ in order to convert from diet-Salpeter to Salpeter IMF.}. We use their orthogonal linear fit considering the stellar mass as the independent variable.  As expected, there is an offset between this classical derivation and our results for the integrated kinematics. However, for the resolved kinematics, which determines $V_{max}$, the offset tends to disappear, at least for the spiral galaxies.

\subsection{FJ Relation.}

Central panels of Figure \ref{fig:scaling_relations} and Figure \ref{fig:SK_V1200} show the FJ distributions including late-type galaxies using the integrated kinematics sample and the spatial resolved one, respectively. A reference FJ relation, derived by \citet{Gallazzi+2006}, has been included for comparison. Our stellar FJ relations show a scatter of $\sim$ 0.16 dex ($\sim$0.14 dex) and $\sim$ 0.17 dex ($\sim$0.10 dex) for gaseous and stellar kinematics, respectively, for the integrated (spatially resolved) subsamples. These dispersions are similar to the ones found by \citetalias{Cortese+2014} ($\sim$ 0.16 dex), but larger than the one reported by \citet{Gallazzi+2006} ($\sim$ 0.07 dex). 

On a parallel situation as the one found for the TF relation, the stellar velocity dispersions and those derived for early-type ones are more in agreement with the FJ relation than the gaseous dispersions and/or those derived for late-type galaxies.

\subsection{$M_\star-S_{K}$ relation.}

Figure \ref{fig:scaling_relations}, right panels, shows the $M_\star$ - $S_{0.5}$ distribution for the integrated kinematics segregated by gas and stellar kinematics (upper panel) and by morphology (lower panel). As a reference the $S_{0.5}$ relations, derived by \citetalias{Cortese+2014} and \citet{Kassin+2007}, have been included, together with the best-fit relation derived with our own data. As in the previous cases, the best-fit parameters for the linear regression have been included in Table \ref{table}. The distribution is clearly tighter than those of the FJ relations, with scatter very similar or lower to the one found for the TF relation ($\sim$0.08 dex).

Figure \ref{fig:SK_V1200}, right panels, shows the same distributions for the resolved kinematics subsample. For this control sample, the scatter decreases significantly to 0.053 dex and 0.052 dex for both the ionized gas and stellar kinematics, respectively. As we mentioned above, the slope, zero-point and scatter of the TF and FJ relations could depend on several factors including (i) the morphology of the galaxies, (ii) the adopted shape for the rotational curve and even (iii) the methodology used to measure both the rotational velocity and/or the velocity dispersion (see Colleen et al., for an example of the effects of the uncertainties). For the $S_{0.5}$ parameter, the dependence on morphology and the described offsets between gaseous and stellar kinematics eventually disappear. Thus, galaxies of any morphology lie on the same scaling relation in agreement with previous studies. 

\citetalias{Cortese+2014} found a good agreement in the slope of the $S_{0.5}$ relation derived using integrated kinematics up to 1$r_e$ for the SAMI dataset with that derived by \citet{Kassin+2007} for a sample of star-forming galaxies, using the maximum rotational velocities. However, they found larger differences in the zero-point of their relations. In our analysis the behavior is similar. The slope remains unchanged between both the integrated and resolved kinematics, with small differences compared with the ones derived by \citet{Cortese+2014} and \citet{Kassin+2007}. However, our best-fit for the total sample (gas + stars) presents a scatter clearly lower than the one found by previous studies, being $\sim$ 0.082 dex for the integrated kinematics and $\sim$ 0.054 dex for the spatial resolved one. The reduction in the scatter combining rotation velocity and velocity dispersion in a single parameter, indicate that together they trace the gravitational potential than each one separately. Actually, this latter value is in agreement with the physical interpretation of \citet[][]{Zaritsky+2008}.

\section{Discussion.}
\label{sec:Disc}

We discuss here the implications of the results listed in the previous section, trying to understand how the uncertainties may affect them and the physical nature of the described relations.

\subsection{Narrowing down the uncertainties.}

A critical challenge giving a physical interpretation to galaxy scaling relations are the uncertainties, because they can potentially modify or erase the dependence between the analyzed properties. We have tried to narrow down their effects by performing the analysis twice. Once using the integrated kinematics, following \citetalias{Cortese+2014}, and then, we improved the accuracy by using a spatial resolved kinematic analysis. This second analysis is performed  at the expenses of the statistics. We consider that this second dataset is best suited to derive a more accurate $S_{0.5}$ relation.

\begin{figure*}
\centering
\epsfig{file=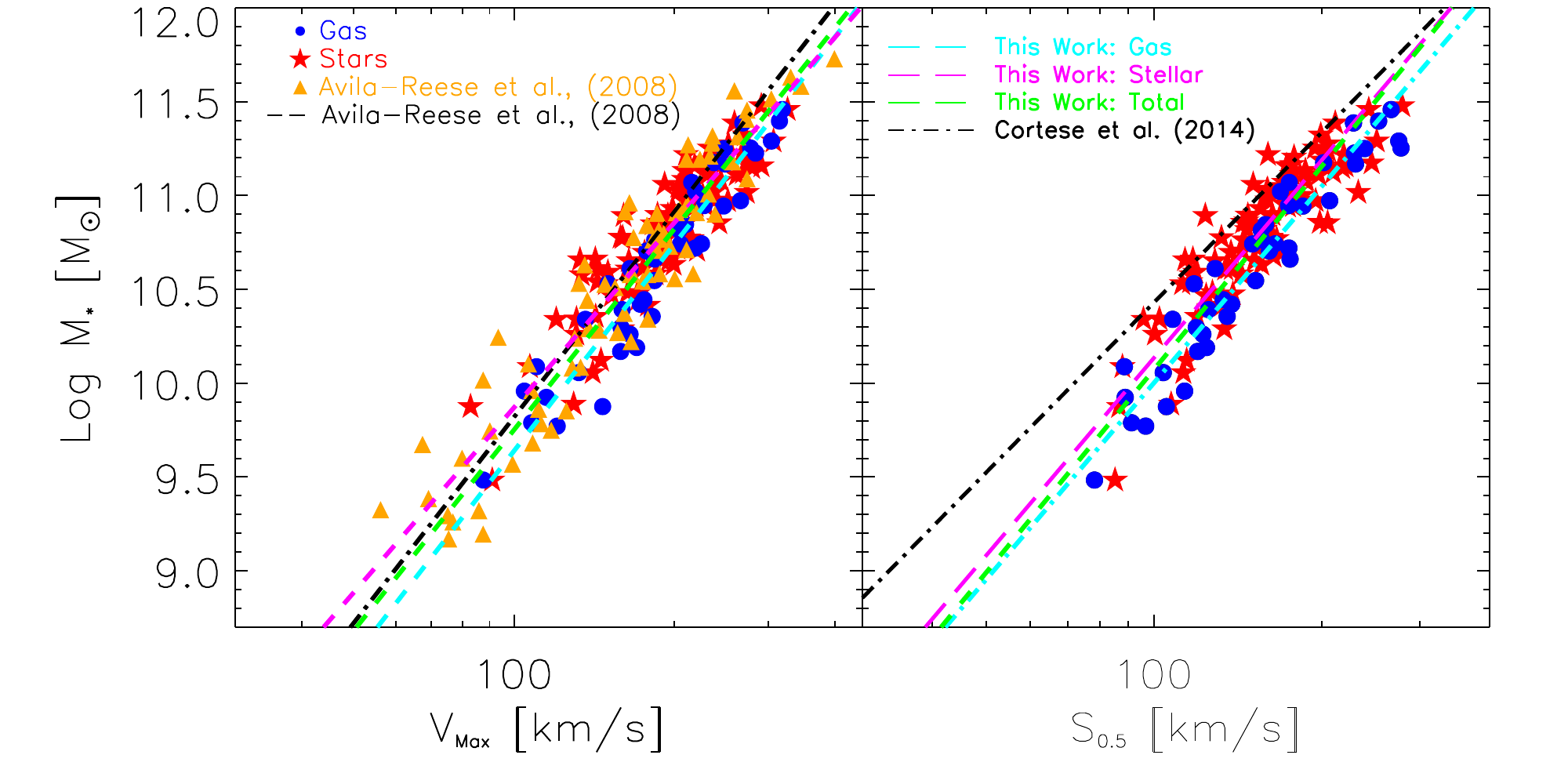,width=15cm, height=8cm}
\caption{Scaling relations for our control subsample. Left and right panels shown the TF and $S_{0.5}$ scaling relations, respectively. Blue and red symbols represent gaseous and stellar kinematics for galaxies with inclinations from $30^{\circ} < i < 70^{\circ}$. Cyan, magenta and green lines are the best-fit for gas, stellar and total (gas + stellar). In the TF relation we recover in great detail the result of the data compilation from \protect\citet[][their masses were corrected to convert to a Salpeter IMF]{Avila-Reese+2008}. It is clear that in galaxies where the random motions are negligible, the $S_{0.5}$ relation tends to be the TF.}
\label{fig:TFySK}
\end{figure*}

Table \ref{table} shows there is a clear improvement in the TF and $S_{0.5}$ relations (in most of the cases) when adopting the spatial resolved kinematics. On the other hand, there is only a mild improvement in the FJ relation (since this relation does not involve rotation velocities). To verify the scatter we tried to reproduce the "classical" TF relation using the spatial resolved kinematics. For doing so, we select only the spiral galaxies and compare their distribution in the $M_\star$-$V_{max}$ diagram with that of a well established comparison sample: the compilation and homogenization presented in \citet{Avila-Reese+2008}. Figure \ref{fig:TFySK}, left panel, shows this comparison. The parameters derived for the TF relation for both subsamples match pretty well, with very good agreement, in particular for the gas kinematics, as shown in Table \ref{table}. Therefore, the spatial resolved kinematic sample seems to be the best one to characterize the scaling relations involving rotation velocities. 

\begin{table*}
\begin{center}
\begin{turn}{90}
\begin{tabular}{lccc}
\hline
\hline
Relation &\hspace{0.5cm} Tully-Fisher & \hspace{0.35cm} Faber-Jackson & \hspace{0.35cm} $S_{0.5}$\\
 &\hspace{0.35cm} \rule{4.0cm}{0.8pt} & \hspace{0.35cm} \rule{4.0cm}{0.8pt}  & \hspace{0.35cm} \rule{4.0cm}{0.8pt}\\
 & \hspace{-2.0cm} \begin{tabular}{lccc} Galaxies & \hspace{0.0cm} scatter & slope & zero-point \end{tabular}
 & \begin{tabular}{lll} scatter & slope & zero-point \end{tabular}
 & \begin{tabular}{lll} scatter & slope & zero-point \end{tabular}\\
\hline
\textbf{Integrated kinematics at $R_e$} & & & \\
Gas &\hspace{-1.0cm} \begin{tabular}{rlll} 223 & \hspace{1.0cm} 0.084 & 0.27 $\pm$ 0.01 & -0.65 $\pm$ 0.12 \end{tabular} & \begin{tabular}{ccc} 0.171 & 0.36 $\pm$ 0.02 & -2.03 $\pm$ 0.27 \end{tabular} & \begin{tabular}{ccc} 0.087 & 0.29 $\pm$ 0.01 & -1.03 $\pm$ 0.12
\end{tabular}\\
Stellar & \hspace{-1.0cm} \begin{tabular}{rlll} 278 & \hspace{1.0cm} 0.200 & 0.16 $\pm$ 0.02 & 0.32 $\pm$ 0.30 \end{tabular} & \begin{tabular}{ccc} 0.160 & 0.31 $\pm$ 0.03 & -1.37 $\pm$ 0.25 \end{tabular} & \begin{tabular}{ccc} 0.075 & 0.26 $\pm$ 0.01 & -0.67 $\pm$ 0.10 \end{tabular}\\
Total & \hspace{-1.0cm} \begin{tabular}{rlll} 501 & \hspace{1.0cm} 0.171 & 0.20 $\pm$ 0.01 & -0.01 $\pm$ 0.18\end{tabular} & \begin{tabular}{ccc} 0.165 & 0.34 $\pm$ 0.02 & -1.71 $\pm$ 0.18\end{tabular} & \begin{tabular}{ccc} 0.082 & 0.27 $\pm$ 0.01 & -0.79 $\pm$ 0.07\end{tabular}\\
\citet{Cortese+2014} & \hspace{-1.0cm} \begin{tabular}{rlll} & \hspace{1.0cm} 0.26 \hspace{1cm} & ----\hspace{1.0cm} & ---- \end{tabular} & \begin{tabular}{ccc} 0.16 \hspace{1cm} & ----\hspace{1cm} & ---- \end{tabular} & \begin{tabular}{ccc} 0.10 & 0.33 $\pm$ 0.01 & -1.41 $\pm$ 0.08 \end{tabular}\\
\citet{Kassin+2007} & \hspace{-1.0cm} \begin{tabular}{rlll} & \hspace{1.0cm} ---- \hspace{1.0cm} & ---- \hspace{1.0cm} & ---- \end{tabular} & \begin{tabular}{ccc} ---- \hspace{1.0cm} & ---- \hspace{1.0cm} & ---- \end{tabular} & \begin{tabular}{ccc} 0.10 & 0.34 $\pm$ 0.05 & 1.89 $\pm$ 0.03\end{tabular}\\
\hline
\textbf{Resolved kinematics, $V_{max}$} & & & \\
Gas & \hspace{-1.0cm} \begin{tabular}{rlll} 59 & \hspace{1.0cm} 0.07 & 0.25 $\pm$ 0.02 & -0.41 $\pm$ 0.17 \end{tabular} & \begin{tabular}{ccc} 0.10 & 0.31 $\pm$ 0.01 & -1.47 $\pm$ 0.19 \end{tabular} & \begin{tabular}{ccc} 0.053 & 0.29 $\pm$ 0.01 & -0.92 $\pm$ 0.13 \end{tabular}\\
Stellar & \hspace{-1.2cm} \begin{tabular}{rlll} 167 & \hspace{1.0cm} 0.24 & -0.10 $\pm$ 0.09 & 3.34 $\pm$ 1.07\end{tabular} & \begin{tabular}{ccc} 0.14 & 0.53 $\pm$ 0.03 & -3.88 $\pm$ 0.34\end{tabular} & \begin{tabular}{ccc} 0.052 & 0.27 $\pm$ 0.01 & -0.72 $\pm$ 0.12 \end{tabular}\\
Total & \hspace{-1.3cm} \begin{tabular}{rlll} 226 & \hspace{1.0cm} 0.22 & 0.08 $\pm$ 0.04 & 1.37 $\pm$ 0.48 \end{tabular} & \begin{tabular}{ccc} 0.13 & 0.44 $\pm$ 0.02 & -2.79 $\pm$ 0.22\end{tabular} & \begin{tabular}{ccc} 0.054 & 0.27 $\pm$ 0.01 & -0.71 $\pm$ 0.11\end{tabular}\\
\hline
\textbf{Only spiral galaxies, $V_{max}$} &  &  &  \\
Gas & \hspace{-1.0cm} \begin{tabular}{rlll} 42 & \hspace{1.0cm} 0.043 & 0.27 $\pm$ 0.01 & -0.63 $\pm$ 0.15 \end{tabular} & \begin{tabular}{ccc} 0.076 & 0.35 $\pm$ 0.02 & -1.84 $\pm$ 0.21\end{tabular} & \begin{tabular}{ccc} 0.043 & 0.29 $\pm$ 0.01 & -0.88 $\pm$ 0.13\end{tabular}\\
Stellar & \hspace{-1.0cm} \begin{tabular}{rlll} 92 & \hspace{1.0cm} 0.053 & 0.30 $\pm$ 0.02 & -1.00 $\pm$ 0.02\end{tabular} & \begin{tabular}{ccc} 0.091 & 0.35 $\pm$ 0.03 & -1.94 $\pm$ 0.34\end{tabular} & \begin{tabular}{ccc}0.052 & 0.28 $\pm$ 0.02 & -0.92 $\pm$ 0.21\end{tabular}\\
Total & \hspace{-1.2cm} \begin{tabular}{rlll} 134 & \hspace{1.0cm} 0.052 & 0.28 $\pm$ 0.01 & -0.73 $\pm$ 0.13\end{tabular} & \begin{tabular}{ccc} 0.098 & 0.33 $\pm$ 0.02 & -1.69 $\pm$ 0.22\end{tabular} & \begin{tabular}{ccc} 0.051 & 0.27 $\pm$ 0.01 & -0.80 $\pm$ 0.13\end{tabular}\\
\citet{Avila-Reese+2008} & \hspace{-1.0cm} \begin{tabular}{rlll} & \hspace{1.2cm} 0.045 & 0.27 $\pm$ 0.01 & -0.69 $\pm$ 0.12\end{tabular} & \begin{tabular}{ccc} ---- \hspace{1cm} & ---- \hspace{1.0cm} & ---- \end{tabular} & \begin{tabular}{ccc} ---- \hspace{1cm} & ---- \hspace{1.0cm} & ---- \end{tabular}\\
\hline

\end{tabular}
\end{turn}
\caption{Orthogonal linear fit parameters to scaling relations. \textbf{Note.} All scatters are estimated from the linear fit as the standard deviation of all residuals, we consider stellar mass, $M_\star$, as independent variable. $log(V,\sigma,S_{0.5})\ =\ a\ +\ blog(M_\star)$. $V,\ \sigma$ and $S_{0.5}$ are given in [km/s],\ $M_\star$ in $M_\odot$.}
\label{table}
\end{center}
\end{table*}

Using this new sub-sample we derive the most precise estimation of the $M_\star$-$S_{0.5}$ relation, shown in Fig. \ref{fig:TFySK}, right panel. The parameters of this relation are listed in Table \ref{table}. The first result emerging from this analysis is that the scatter is of the order of the $S_{0.5}$ relation found for the complete resolved kinematics ($\sim$0.05 dex). Therefore, to select a better sub-sample in terms of the TF relation does not seem to affect the result. In other words, the inclusion of early-type galaxies affects the TF relation, but it does not affect the $S_{0.5}$ one. Another interesting result is that the scatter in this relation is very similar to that of the TF relation for the same subsample. Therefore, the inclusion of the effects of random motions does not increase the scatter, even for galaxies clearly supported by rotation.

Finally, the slope and zero-point of the $S_{0.5}$ relations found for (i) this particular sub-sample of galaxies that reproduces better the TF relation, (ii) the complete resolved kinematics sample, and (iii) the integrated kinematics sample, when considering both the gaseous and stellar kinematics, agree with each other. Thus, only the precision is increased by performing a detailed resolved kinematics for a TF-compatible subsample, but the general trends are the same.
The result of this test suggests that our analysis is not dominated by velocity uncertainties and the early tight correlation presented by \citetalias{Cortese+2014} and in this paper is real and not the result of the poorly constrained in velocity for dispersion dominated systems.

\subsection{$S_K$ as a proxy of the dynamical mass.}

The observed kinematics of a galaxy often is used to infer the total (dynamical) mass enclosed at different radii \citep[e.g.][]{Persic1988,Zavala+2003,Courteau2014,Ouellette+2017}. Assuming that the $M_\star - S_{0.5}$ scaling relation is a consequence of a more physical relation between the dynamical mass and the stellar mass in the inner regions, we suppose that the $S_{0.5}$ parameter traces the dynamical mass as follow:

\begin{equation}
M_{dyn} \propto S_{0.5}^2 \Rightarrow M_{dyn} = \eta \frac{r_r S_{0.5}^2}{G}=\eta \frac{r_{r}(0.5V_{rot}^2 + \sigma^2)}{G},
\label{factor}
\end{equation}
where $r_r$ is a characteristic radius of the galaxy, $G$ the gravitational constant and $\eta$ is a structural coefficient which encapsulate information of the shape of the galaxy, projection effects, dynamical structure, etc., in fact it can be included into the $K$ coefficient of the $S_{0.5}^2$ parameter, however it is useful to introduce $\eta$ in order to compare with former studies. Dynamical models like Jeans Anisotropic Models \citep[JAM's,][]{Cappellari2008} or Schwarzschild \citep{Schwarzschild1979} are considered the state-of-the-art inferences of galaxies mass distribution includying dynamical enclosed mass. \citet{Cappellari2006} calibrated eq. \ref{factor} for a sample of early-type galaxies from the SAURON project \citep{Bacon2001} using the velocity dispersion instead the $S_{0.5}$ parameter in combination with Schwarzschild dynamical models. They found that the dynamical mass within the effective radius can be robustly recovered using a coefficient $\eta \approx 2.5$, which varies little from galaxy to galaxy.

\begin{figure}
\centering
\epsfig{file=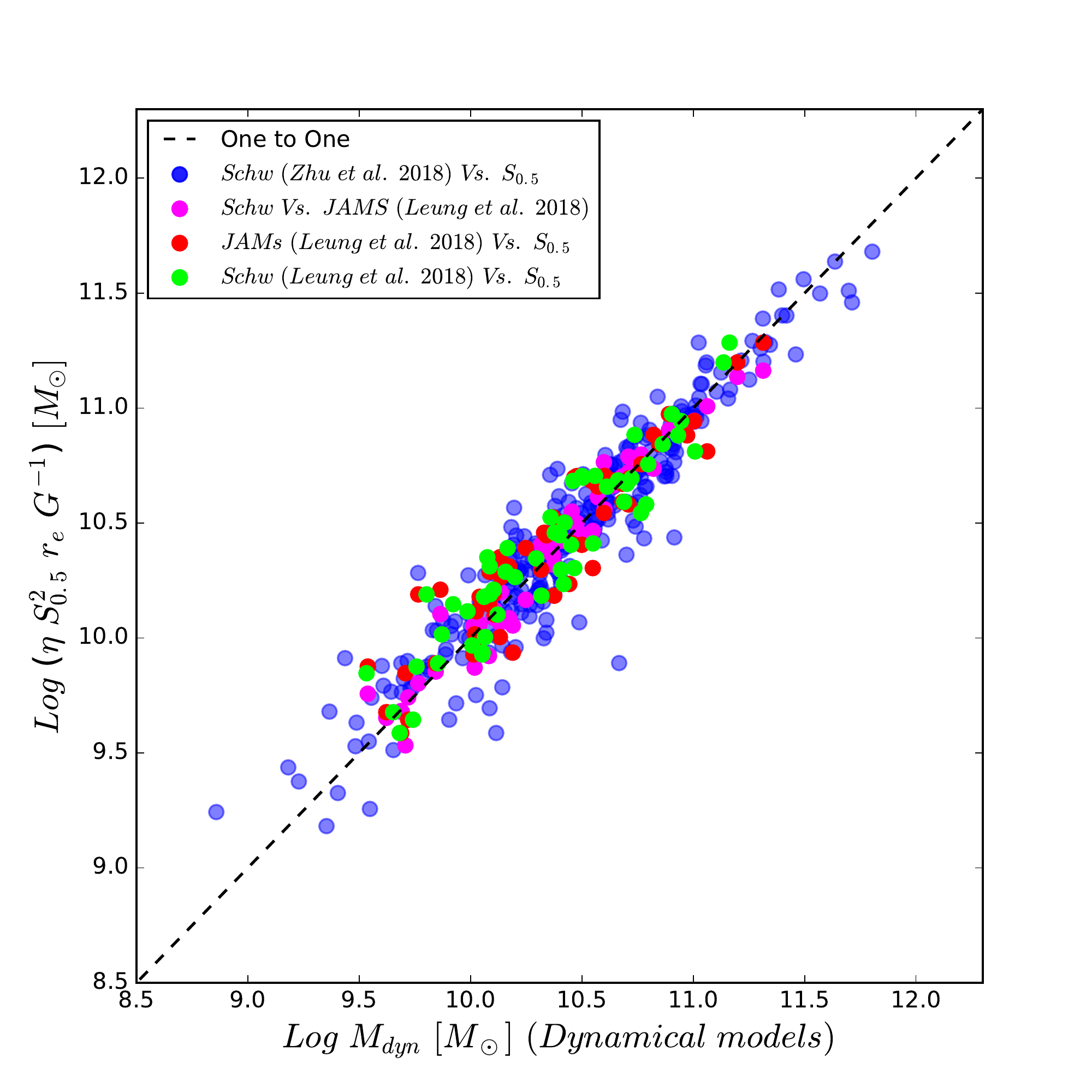,width=9cm, height=9cm}
\caption{One to one relation between dynamical masses inferred from dynamical models and kinematic parameter $S_{0.5}$. Blue symbols are the comparison between the Schwarzschild models by \protect\citet{Zhu+2017} with our estimations. Red and green symbols are the comparison between JAMs and Schwarzschild models by \protect\citet{Leung2018} with our estimations, respectively. Both comparisons shown a scatter of $\sim 0.15\ dex$. Magenta symbols are the comparison between Schwarzschild and JAMs estimations with a scatter of 0.08dex.}
\label{fig:Masses}
\end{figure}

\begin{figure*}
\centering
\epsfig{file=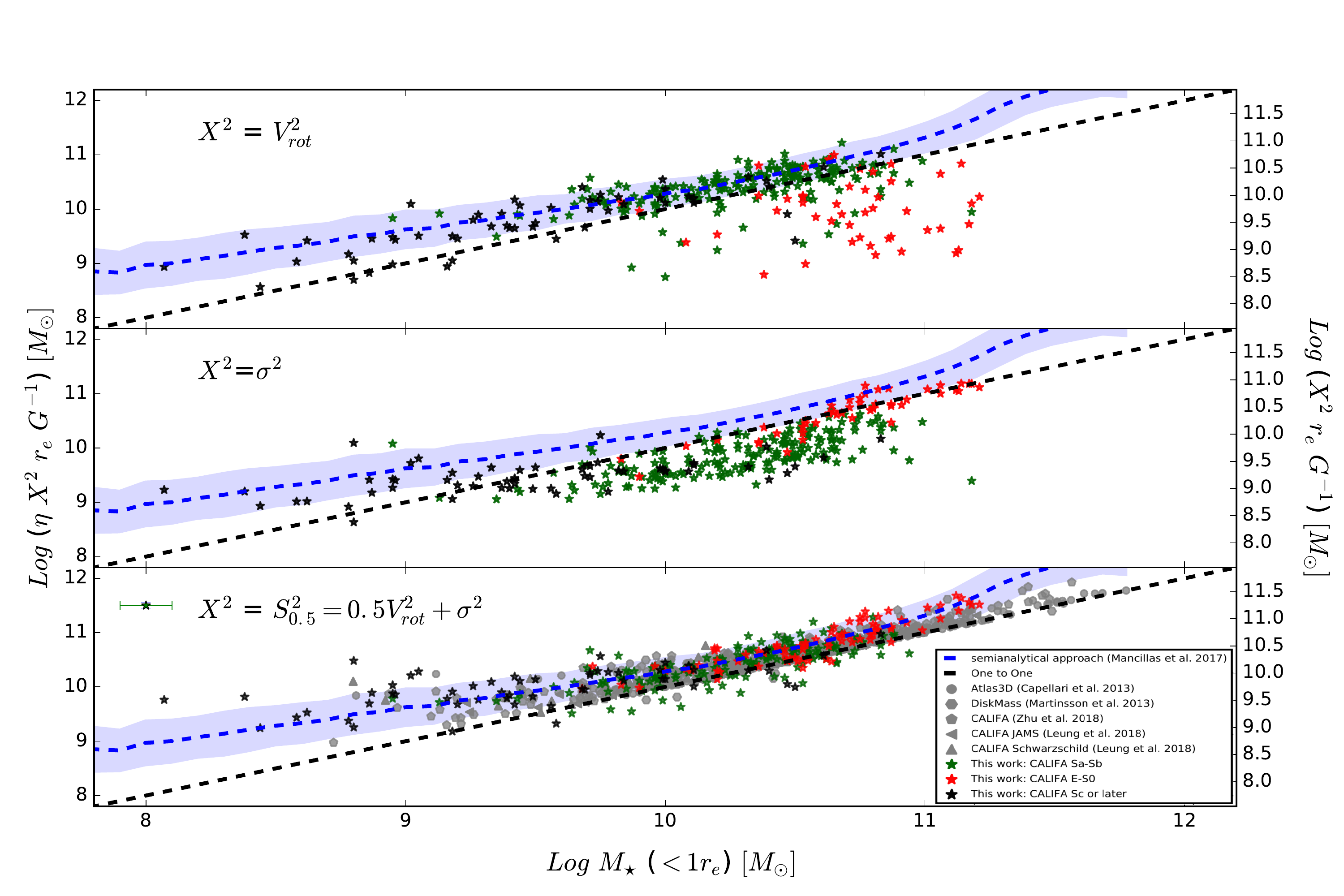,width=16cm, height=14cm}
\caption{Accuracy of the M$_{dyn}$-M$_\star$ relations based on the $S_{0.5}$ parameter. In top and medium panels we assume that galaxies are rotation or velocity dispersion dominated to estimate the dynamical mass within the effective radius. Red, green and black star symbols represent our CALIFA sample, whereas gray symbols are from the literature compilation. The $S_{0.5}$ dynamical mass estimations perform better than the ones based either only on rotation or dispersion. In the bottom panel we used the $S_{0.5}$ parameter to estimate the dynamical mass using equation (\ref{factor}) and compare them with theoretical predictions based on detailed dynamical models. All the estimated M$_{dyn}$-M$_\star$ relations are comparable and consistent with observations within the uncertainties. As a reference we also show the semi-empirical predictions of \protect\citet{Mancillas+2017} (blue shaded region; see text) which use $\eta=1$ and are also consistent with our estimations.}
\label{fig:Mdyn}
\end{figure*}

\citet{Leung2018} performed a detailed comparison of JAMs and Schwarzschild models for 54 of the CALIFA galaxies included here. We use these dynamical masses, $M_{dyn}^{JAMs}$, to calibrate the eq. \ref{factor} based on the $S_{0.5}$ parameter. We found that the enclosed dynamical mass within the effective radius (i.e., using r$_e$ as the characteristic radius in eq. \ref{factor}) can be robustly recovered using a single coefficient $\eta \approx 1.8$ for all the galaxies, with a narrow dispersion of $0.15\ dex$. To validate that calibration we compare the estimated dynamical masses by the eq. \ref{factor} with those derived using dynamical models for a sample of 300 galaxies analyzed by \citet{Zhu+2017,Zhu2+2018}, together with the ones by \citet{Leung2018}. Figure \ref{fig:Masses} shows the comparison between the different estimations of the dynamical masses. As expected, the agreement between the values derived using JAMs and Schwarzschild dynamical models for the galaxies studied by \citet{Leung2018} agree one each other with a low scatter of $0.08\ dex$. Interestingly, we still find a very good agreement using $\eta = 1.8$  between our $S_{0.5}$ derived dynamical masses and sophisticated dynamical mass estimations, with a scatter of $\sim 0.15\,dex$.  We may wonder why is the  $S_{0.5}$ parameter such a good mass tracer.  This is remarkable in view that,  we do not systematically study IMF effects \citep[e.g.][]{nacho15} and kinematic anisotropy \citep[e.g.][]{Zhu+2017}. The enclosed mass within $r_e$ is an integrated quantity weakly sensitive to the specific mass and shape density profile, a similar discussion has been presented by \citet{Wolf+2010} for dwarf spheroidal galaxies, only in such grounds the $S_{0.5}$ is a competitive $M_{dyn}$ proxy. 

\subsection{The Dynamical-to-stellar mass relation.}

We explore the literature in order to compile the most recent state-of-the-art derivations of the dynamical mass in the central regions of galaxies using dynamical models. \citet{Cappellari+2013} estimated the dynamical mass within the effective radius for 258 early-type galaxies from the $ATLAS^{3D}$ project \citep{Cappellari+2011} using the JAM's dynamical models and compared it with the stellar masses. \citet{Martinsson+2013} performed a similar study for 24 late-type galaxies extracted from the DiskMass survey \citep{bershady10}. \citet{Zhu+2017,Zhu2+2018} constructed orbit-superposition Schwarzschild models at different radii that simultaneously fit the observed surface brightness and stellar kinematics for 300 galaxies included in the CALIFA-V1200 resolution subsample studied here. In \citet{Zhu2+2018} they constrained the stellar orbit distribution and found that a fraction of stars are within a plane with unperturbed orbits tracing the rotation velocity, while others are out of the plane with perturbed orbits tracing the velocity dispersion. This result implies that the kinematics in galaxies is more complex that just rotation or velocity dispersion: both components are present in all types of galaxies and should be considered to trace the potential. Figure \ref{fig:Mdyn}, bottom panel, presents the comparison between the distributions of dynamical masses along the stellar ones between this compilation of data extracted from the literature and those ones derived using the eq. \ref{factor} within the effective radius, based on the $S_{0.5}$ parameter with $\eta = 1.8$. In addition we present the dynamical masses derived if we consider only the rotational velocities or the velocity dispersions.\footnote{Corrections for different adopted values for the effective radius and IMFs offset have been considered when required.} All these dynamical masses, derived at $r_e$ are listed in Table \ref{tab:Gal}.
We observe a clear offset and a large scatter between our dynamical masses and those derived using detailed models when we use only rotation velocity (mostly for ellipticals) or velocity dispersion (mostly for spirals). However, when we use the dynamical mass proxy based on the $S_{0.5}$ parameter, the distribution along the stellar mass is in agreement with the results extracted from the literature. Thus, it seems that the $S_{0.5}$ parameter is indeed a good proxy for calculating the dynamical mass.

Our distribution of $M_{dyn}-M_\star$ follows a linear and nearly one-to-one relation for masses in the range $3\times 10^{9}\la M_\star\ [M_{\odot}] \la 5\times 10^{10}$. The fact that for some galaxies (both from our sample and from other works), the stellar mass seems to be higher than the dynamical one shows the presence of several systematical uncertainties both in the stellar and dynamical mass determinations.
Within these uncertainties, what we learn from Fig. \ref{fig:Mdyn} is that in the mentioned above mass range luminous matter strongly dominates within $1r_e$.  Below $\sim 3\times 10^{9} M_\odot$ there is a clear deviation, with galaxies showing larger dynamical masses than their stellar masses, which indicates that in the low-mass regime galaxies are more dark-matter dominated as less massive they are, even within the effective radius. In the high-mass end, there is some weak evidence of a deviation, with the few E/S0 galaxies at these masses showing again larger dynamical masses than their stellar masses. This difference could be due to more bottom-heavy IMF \citep[e.g.][]{lyubenova16} and/or due to the contribution of dark matter.

Both our data and literature collected ones show similar trends.
Indeed, this result is predicted by different theoretical studies, including hydrodynamical cosmological simulations \citep[e.g.][]{Oman+MNRAS}  from the Evolution and Assembly of GaLaxies and their Environments (EAGLE) project \citep[][]{Schaye+2015,Crain+2015}, and semi-empirical modeling approaches \citep[e.g.,][]{Mancillas+2017}. We include the latter theoretical predictions for comparison in Figure \ref{fig:Mdyn}. In \citet[][]{Mancillas+2017}, a population of galaxies with bulge-to-disk mass ratios lower than $\sim 0.7$ was generated by loading the bulge/disc systems into $\Lambda$CDM halos, taking into account the adiabatic contraction of the inner halo by the baryons. The modeled population reproduces well the TF relation, radius-mass, $B/T$-mass, and gas-to-stellar mass relations, and by construction follows the stellar-to-halo ($M_\star$-$M_{vir}$) relation constrained from a semi-empirical approach for blue galaxies \citep{Rodriguez-Puebla+2015}. The predicted inner mass distributions, in particular the stellar-to-dynamical masses within 1$r_e$, inherit partially the shape of the latter relation, which bends to lower $M_\star/M_{vir}$ ratios both at lower and higher masses. This explains the bends seen for the predictions in Fig. \ref{fig:Mdyn} (dashed blue line and shadow region around it). It is encouraging that our observational inferences based on the $S_{0.5}$ parameter agrees with these predictions, showing the possibility to attain a connection between the inner galaxy dynamics of the local galaxy population and the properties of the cosmological dark matter halos. 






\section{Conclusions}
\label{sec:Concl}
Originally the $S_{K}$ parameter was introduced as a tool to deal with galaxies difficult to classify or with high amount of velocity dispersion like clumpy high redshift galaxies. The remarkable reduction of scatter in the $S_{0.5}$ relationship compared with TF and FJ relations found by previous studies \citep[e.g][]{Cortese+2014} and confirmed with higher accuracy by our study, points toward a more complex internal kinematics in galaxies even in the local Universe: non-circular motions in disk galaxies and some amount of rotation in elliptical galaxies.

In summary, we demonstrate that (i) the $M_\star-S_{0.5}$ is a tighter correlation than the TF relation or the FJ relation when galaxies of all morphological types are considered, and (ii) this relation is a consequence of $S_{0.5}$ being a proxy of the dynamical mass and the relation between this later parameter with the stellar mass. Finally, we propose a simple but competitive procedure to estimate the dynamical mass in galaxies, easier to apply to massive surveys than more detailed analysis, although with lower precision.
 
\section*{Acknowledgements}

We thank the support by CONACYT grant CB-285080. OV and EA acknowledge support from the PAPIIT grant  IN112518. S.F.S. thank PAPIIT-DGAPA-IA101217 (UNAM) project. We would like to thank Damian Mast for his valuable job observing the CALIFA galaxies. Many thanks to Gigi Y. C. Leung from the Max Planck Institute for Astronomy for providing us the stellar and dynamical masses from dynamical models for the 54 galaxies from her study. GvdV acknowledges funding from the European Research Council (ERC) under the European Union's Horizon 2020 research and innovation programme under grant agreement No 724857 (Consolidator Grant ArcheoDyn).

This study  uses data provided by the Calar Alto Legacy
Integral Field Area (CALIFA) survey (http://califa.caha.es/).

CALIFA is the first legacy survey performed at Calar Alto. The
CALIFA collaboration would like to thank the IAA-CSIC and MPIA-MPG as
major partners of the observatory, and CAHA itself, for the unique
access to telescope time and support in manpower and infrastructures.
The CALIFA collaboration also thanks the CAHA staff for the dedication
to this project.

Based on observations collected at the Centro Astron\'omico Hispano
Alem\'an (CAHA) at Calar Alto, operated jointly by the
Max-Planck-Institut f\"ur Astronomie and the Instituto de Astrof\'\i sica de
Andaluc\'\i a  (CSIC).




\bibliographystyle{mnras}
\bibliography{refs,CALIFAI} 




\appendix
\section{Stellar and dynamical masses.}
\label{Appendix: AppendixA}

\restylefloat{table}

Dynamical masses were estimated within $1r_e$ using Eq. \ref{factor}

\begin{table}[h]
\caption{Stellar masses and dynamical masses within the effective radius.}
\begin{tabular}{lccl} 
\textbf{Name} & \textbf{$M_\star$} & \textbf{$M_{dyn}$} & \textbf{$r_e$}\\
       & $[M_\odot]$ & $[M_\odot]$ & [arcsec]\\
      (1) & (2) & (3) & (4) \\
      \hline
IC5376 & 10.16$\pm$0.10 & 10.53$\pm$0.04 & 11.62\\
NGC0036 & 10.76$\pm$0.09 & 10.82$\pm$0.02 & 19.34\\
UGC00148 & 9.71$\pm$0.09 & 10.26$\pm$0.06 & 13.54\\
MCG-02-02-030 & 10.00$\pm$0.09 & 10.25$\pm$0.03 & 13.86\\
UGC00005 & 10.62$\pm$0.09 & 10.78$\pm$0.01 & 14.45\\
NGC7819 & 10.00$\pm$0.08 & 10.14$\pm$0.03 & 15.02\\
UGC00029 & 10.93$\pm$0.10 & 11.19$\pm$0.04 & 12.79\\
IC1528 & 10.04$\pm$0.09 & 10.16$\pm$0.03 & 16.95\\
NGC7824 & 10.64$\pm$0.09 & 10.75$\pm$0.12 & 9.64\\
UGC00312 & 9.75$\pm$0.09 & 10.57$\pm$0.07 & 13.28\\
MCG-02-02-040 & 9.44$\pm$0.09 & 10.11$\pm$0.05 & 11.62\\
UGC00335NED02 & 10.43$\pm$0.10 & 10.72$\pm$0.04 & 16.64\\
\label{tab:Gal}
\end{tabular}\\
\raggedright Notes. \textbf{Col. (1):} Galaxy name. \textbf{Col. (2):} Stellar mass within $r_e$ estimated from PIPE3D. \textbf{Col. (3):} Dynamical mass estimated from the kinematic parameter $S_{0.5}$. \textbf{Col. (4):} Effective radius $r_e$.
\end{table}

\begin{table}[t!]
\renewcommand\thetable{A1}
      \caption{continuation}
    \label{tab:Gal}
\begin{tabular}{lccl} 
\textbf{Name} & \textbf{$M_\star$} & \textbf{$M_{dyn}$} & \textbf{$r_e$}\\
       & $[M_\odot]$ & $[M_\odot]$ & [arcsec]\\
      (1) & (2) & (3) & (4) \\
      \hline
NGC0216 & 8.78$\pm$0.09 & 9.38$\pm$0.08 & 13.22\\
NGC0214 & 10.66$\pm$0.09 & 10.46$\pm$0.04 & 14.88\\
NGC0217 & 10.37$\pm$0.09 & 10.83$\pm$0.01 & 20.44\\
NGC0237 & 10.11$\pm$0.09 & 10.01$\pm$0.05 & 11.05\\
NGC0234 & 10.50$\pm$0.08 & 10.00$\pm$0.04 & 17.36\\
MCG-02-03-015 & 10.94$\pm$0.10 & 10.62$\pm$0.02 & 11.56\\
NGC0257 & 10.61$\pm$0.09 & 10.57$\pm$0.02 & 15.10\\
NGC0364 & 10.36$\pm$0.09 & 10.66$\pm$0.01 & 9.04\\
NGC0429 & 10.09$\pm$0.09 & 10.38$\pm$0.04 & 7.14\\
IC1652 & 10.21$\pm$0.09 & 10.27$\pm$0.05 & 10.62\\
NGC0447 & 10.55$\pm$0.09 & 10.71$\pm$0.05 & 18.56\\
NGC0444 & 9.71$\pm$0.10 & 10.22$\pm$0.04 & 17.37\\
UGC00809 & 9.02$\pm$0.08 & 10.21$\pm$0.03 & 11.01\\
UGC00841 & 9.73$\pm$0.11 & 10.28$\pm$0.05 & 13.73\\
NGC0477 & 10.39$\pm$0.09 & 10.54$\pm$0.03 & 18.58\\
IC1683 & 10.31$\pm$0.09 & 10.18$\pm$0.03 & 9.97\\
NGC0499 & 10.64$\pm$0.08 & 11.01$\pm$0.03 & 13.16\\
NGC0496 & 10.40$\pm$0.11 & 10.30$\pm$0.03 & 16.47\\
NGC0504 & 9.72$\pm$0.10 & 10.38$\pm$0.03 & 8.53\\
NGC0517 & 10.13$\pm$0.10 & 10.40$\pm$0.03 & 7.52\\
UGC00987 & 10.33$\pm$0.09 & 10.40$\pm$0.03 & 10.95\\
NGC0528 & 10.46$\pm$0.10 & 10.59$\pm$0.02 & 9.01\\
NGC0529 & 10.63$\pm$0.09 & 10.84$\pm$0.05 & 11.75\\
NGC0551 & 10.33$\pm$0.10 & 10.46$\pm$0.04 & 14.37\\
UGC01057 & 10.11$\pm$0.10 & 10.31$\pm$0.03 & 11.00\\
UGC01271 & 10.28$\pm$0.10 & 10.47$\pm$0.04 & 8.17\\
NGC0681 & 9.99$\pm$0.08 & 10.28$\pm$0.04 & 23.63\\
NGC0741 & 11.17$\pm$0.09 & 11.56$\pm$0.02 & 25.68\\
NGC0755 & 9.18$\pm$0.08 & 9.91$\pm$0.09 & 19.11\\
IC1755 & 10.43$\pm$0.09 & 10.95$\pm$0.01 & 13.50\\
NGC0768 & 10.39$\pm$0.08 & 10.67$\pm$0.02 & 15.59\\
NGC0774 & 10.50$\pm$0.10 & 10.54$\pm$0.03 & 13.31\\
NGC0776 & 10.53$\pm$0.08 & 10.07$\pm$0.04 & 13.28\\
NGC0781 & 11.18$\pm$0.09 & 10.07$\pm$0.04 & 8.99\\
NGC0810 & 10.77$\pm$0.10 & 11.39$\pm$0.02 & 13.56\\
NGC0825 & 8.95$\pm$0.10 & 9.79$\pm$0.03 & 2.02\\
NGC0932 & 10.64$\pm$0.09 & 10.47$\pm$0.04 & 16.43\\
NGC1056 & 9.87$\pm$0.09 & 9.55$\pm$0.07 & 7.90\\
NGC1060 & 11.13$\pm$0.09 & 11.40$\pm$0.03 & 20.46\\
UGC02222 & 10.51$\pm$0.10 & 10.47$\pm$0.03 & 8.36\\
UGC02229 & 10.68$\pm$0.09 & 10.96$\pm$0.02 & 11.77\\
NGC1093 & 10.23$\pm$0.08 & 10.31$\pm$0.02 & 8.50\\
UGC02403 & 10.21$\pm$0.09 & 10.24$\pm$0.02 & 11.71\\
NGC1167 & 10.99$\pm$0.09 & 11.04$\pm$0.01 & 21.55\\
NGC1349 & 10.87$\pm$0.09 & 10.83$\pm$0.02 & 14.13\\
NGC1542 & 9.99$\pm$0.10 & 10.24$\pm$0.03 & 9.53\\
NGC1645 & 10.43$\pm$0.10 & 10.76$\pm$0.01 & 14.09\\
UGC03151 & 10.41$\pm$0.10 & 10.56$\pm$0.03 & 15.50\\
NGC1677 & 9.20$\pm$0.08 & 9.68$\pm$0.10 & 8.59\\
IC2101 & 9.82$\pm$0.10 & 10.26$\pm$0.05 & 14.10\\
UGC03253 & 10.07$\pm$0.09 & 10.31$\pm$0.02 & 12.67\\
NGC2253 & 9.79$\pm$0.10 & 9.59$\pm$0.03 & 4.08\\
UGC03539 & 9.26$\pm$0.08 & 10.02$\pm$0.04 & 13.67\\
NGC2347 & 10.50$\pm$0.09 & 10.58$\pm$0.03 & 13.78\\
UGC03899 & 8.95$\pm$0.10 & 9.88$\pm$0.08 & 9.55\\
UGC00036 & 10.45$\pm$0.09 & 10.71$\pm$0.01 & 10.05\\
NGC0001 & 10.42$\pm$0.09 & 10.19$\pm$0.04 & 9.18\\
NGC0023 & 10.83$\pm$0.08 & 10.59$\pm$0.06 & 10.78\\
\label{tab:Gal}
\end{tabular}
\end{table}

\begin{table}[t!]
\renewcommand\thetable{A1}
      \caption{continuation}
    \label{tab:Gal}
\begin{tabular}{lccl} 
\textbf{Name} & \textbf{$M_\star$} & \textbf{$M_{dyn}$} & \textbf{$r_e$}\\
       & $[M_\odot]$ & $[M_\odot]$ & [arcsec]\\
      (1) & (2) & (3) & (4) \\
      \hline
NGC2410 & 10.49$\pm$0.09 & 10.76$\pm$0.02 & 17.91\\
UGC03944 & 9.57$\pm$0.12 & 10.00$\pm$0.03 & 11.79\\
UGC03969 & 10.34$\pm$0.11 & 10.70$\pm$0.01 & 11.16\\
UGC03995 & 10.64$\pm$0.09 & 10.74$\pm$0.02 & 21.78\\
NGC2449 & 10.30$\pm$0.09 & 10.58$\pm$0.01 & 12.86\\
UGC04029 & 10.09$\pm$0.08 & 10.39$\pm$0.01 & 14.97\\
IC0480 & 9.42$\pm$0.10 & 10.19$\pm$0.02 & 11.49\\
NGC2476 & 10.36$\pm$0.11 & 10.37$\pm$0.05 & 7.99\\
NGC2480 & 8.86$\pm$0.10 & 9.69$\pm$0.10 & 10.82\\
NGC2481 & 9.68$\pm$0.10 & 9.97$\pm$0.03 & 7.54\\
NGC2486 & 10.43$\pm$0.09 & 10.51$\pm$0.03 & 12.96\\
NGC2487 & 10.51$\pm$0.08 & 10.36$\pm$0.04 & 18.81\\
UGC04132 & 10.40$\pm$0.10 & 10.70$\pm$0.02 & 13.18\\
UGC04145 & 10.01$\pm$0.11 & 10.34$\pm$0.03 & 7.93\\
NGC2513 & 10.71$\pm$0.08 & 11.21$\pm$0.02 & 19.23\\
UGC04197 & 9.82$\pm$0.10 & 10.57$\pm$0.03 & 14.93\\
NGC2540 & 10.31$\pm$0.10 & 10.48$\pm$0.02 & 15.42\\
UGC04280 & 9.76$\pm$0.09 & 10.11$\pm$0.05 & 11.18\\
IC2247 & 10.30$\pm$0.09 & 10.53$\pm$0.02 & 16.30\\
UGC04308 & 9.99$\pm$0.08 & 10.15$\pm$0.03 & 21.43\\
NGC2553 & 10.21$\pm$0.09 & 10.52$\pm$0.03 & 8.52\\
NGC2554 & 10.79$\pm$0.09 & 10.87$\pm$0.01 & 17.50\\
NGC2592 & 9.83$\pm$0.10 & 10.04$\pm$0.01 & 7.62\\
NGC2604 & 9.28$\pm$0.10 & 9.77$\pm$0.08 & 20.19\\
NGC2639 & 10.41$\pm$0.08 & 10.70$\pm$0.01 & 13.36\\
UGC04722 & 8.07$\pm$0.12 & 9.76$\pm$0.05 & 17.59\\
NGC2730 & 9.68$\pm$0.08 & 9.96$\pm$0.03 & 14.56\\
NGC2880 & 9.90$\pm$0.08 & 9.98$\pm$0.01 & 13.71\\
IC2487 & 10.00$\pm$0.10 & 10.45$\pm$0.02 & 16.77\\
IC0540 & 9.35$\pm$0.10 & 9.74$\pm$0.03 & 14.88\\
NGC2906 & 9.94$\pm$0.08 & 10.10$\pm$0.02 & 15.23\\
NGC2916 & 10.40$\pm$0.08 & 10.55$\pm$0.02 & 20.60\\
UGC05108 & 10.48$\pm$0.10 & 10.81$\pm$0.03 & 9.58\\
NGC2918 & 10.71$\pm$0.10 & 10.93$\pm$0.02 & 9.32\\
UGC05113 & 10.19$\pm$0.10 & 10.71$\pm$0.02 & 8.91\\
NGC3106 & 10.83$\pm$0.08 & 10.77$\pm$0.03 & 17.30\\
NGC3057 & 8.80$\pm$0.09 & 9.26$\pm$0.09 & 18.08\\
UGC05498NED01 & 9.71$\pm$0.11 & 10.67$\pm$0.02 & 10.52\\
NGC3158 & 11.14$\pm$0.10 & 11.64$\pm$0.03 & 22.29\\
NGC3160 & 10.28$\pm$0.10 & 10.76$\pm$0.01 & 12.77\\
UGC05598 & 9.84$\pm$0.10 & 10.15$\pm$0.04 & 11.40\\
NGC3300 & 10.10$\pm$0.10 & 10.30$\pm$0.01 & 13.31\\
NGC3303 & 10.63$\pm$0.10 & 10.75$\pm$0.03 & 9.24\\
UGC05771 & 10.54$\pm$0.10 & 10.81$\pm$0.03 & 8.01\\
NGC3381 & 9.18$\pm$0.07 & 9.18$\pm$0.10 & 14.84\\
UGC05990 & 8.44$\pm$0.12 & 9.24$\pm$0.09 & 9.36\\
UGC06036 & 10.32$\pm$0.10 & 10.94$\pm$0.02 & 11.16\\
IC0674 & 10.53$\pm$0.09 & 10.84$\pm$0.02 & 11.48\\
UGC06312 & 10.55$\pm$0.09 & 10.80$\pm$0.02 & 12.77\\
NGC3615 & 10.87$\pm$0.09 & 11.10$\pm$0.03 & 10.85\\
NGC3687 & 9.99$\pm$0.08 & 9.75$\pm$0.04 & 15.42\\
NGC3811 & 10.16$\pm$0.09 & 10.08$\pm$0.04 & 14.71\\
NGC3815 & 9.90$\pm$0.08 & 10.12$\pm$0.03 & 8.81\\
NGC3994 & 10.09$\pm$0.10 & 10.27$\pm$0.05 & 7.14\\
NGC4003 & 10.48$\pm$0.09 & 10.52$\pm$0.05 & 9.41\\
UGC07012 & 9.39$\pm$0.08 & 9.79$\pm$0.10 & 11.88\\
NGC4047 & 10.34$\pm$0.09 & 10.41$\pm$0.11 & 14.79\\
\label{tab:Gal}
\end{tabular}
\end{table}

\begin{table}[t!]
\renewcommand\thetable{A1}
      \caption{continuation}
    \label{tab:Gal}
\begin{tabular}{lccl} 
\textbf{Name} & \textbf{$M_\star$} & \textbf{$M_{dyn}$} & \textbf{$r_e$}\\
       & $[M_\odot]$ & $[M_\odot]$ & [arcsec]\\
      (1) & (2) & (3) & (4) \\
      \hline
UGC07145 & 9.96$\pm$0.10 & 10.38$\pm$0.02 & 11.75\\
NGC4149 & 9.70$\pm$0.10 & 10.35$\pm$0.04 & 11.48\\
NGC4185 & 10.20$\pm$0.08 & 10.51$\pm$0.02 & 22.60\\
NGC4210 & 9.85$\pm$0.10 & 9.97$\pm$0.02 & 16.92\\
NGC4470 & 9.58$\pm$0.09 & 9.33$\pm$0.07 & 11.54\\
NGC4644 & 10.11$\pm$0.09 & 10.40$\pm$0.02 & 14.27\\
NGC4711 & 9.97$\pm$0.08 & 10.18$\pm$0.02 & 12.28\\
NGC4816 & 10.75$\pm$0.09 & 11.29$\pm$0.03 & 20.36\\
NGC4841A & 10.82$\pm$0.09 & 11.29$\pm$0.05 & 13.68\\
NGC4874 & 11.12$\pm$0.09 & 11.68$\pm$0.02 & 38.42\\
UGC08107 & 10.80$\pm$0.09 & 11.29$\pm$0.03 & 17.66\\
NGC4956 & 10.53$\pm$0.09 & 10.35$\pm$0.03 & 8.68\\
NGC4961 & 9.42$\pm$0.09 & 9.68$\pm$0.06 & 9.74\\
UGC08231 & 9.05$\pm$0.11 & 10.28$\pm$0.06 & 16.86\\
UGC08234 & 10.62$\pm$0.10 & 10.51$\pm$0.02 & 5.53\\
NGC5000 & 10.22$\pm$0.09 & 10.02$\pm$0.04 & 10.18\\
NGC5016 & 9.98$\pm$0.10 & 10.00$\pm$0.03 & 15.29\\
NGC5029 & 10.86$\pm$0.10 & 11.40$\pm$0.03 & 15.45\\
NGC5056 & 10.32$\pm$0.08 & 10.30$\pm$0.05 & 13.77\\
NGC5205 & 9.43$\pm$0.09 & 9.78$\pm$0.04 & 16.41\\
NGC5216 & 10.08$\pm$0.08 & 10.33$\pm$0.06 & 15.28\\
NGC5218 & 10.15$\pm$0.09 & 10.23$\pm$0.01 & 12.30\\
UGC08733 & 8.96$\pm$0.10 & 9.86$\pm$0.04 & 19.90\\
IC0944 & 10.46$\pm$0.11 & 10.84$\pm$0.02 & 9.80\\
UGC08778 & 9.65$\pm$0.10 & 10.08$\pm$0.04 & 11.90\\
UGC08781 & 10.59$\pm$0.08 & 10.70$\pm$0.04 & 12.01\\
NGC5378 & 10.04$\pm$0.08 & 10.22$\pm$0.03 & 19.29\\
NGC5406 & 10.46$\pm$0.08 & 10.70$\pm$0.01 & 14.93\\
NGC5480 & 9.56$\pm$0.09 & 9.64$\pm$0.06 & 17.41\\
NGC5485 & 10.20$\pm$0.08 & 10.48$\pm$0.02 & 21.81\\
UGC09067 & 10.51$\pm$0.09 & 10.69$\pm$0.03 & 11.26\\
NGC5520 & 9.63$\pm$0.10 & 9.89$\pm$0.03 & 11.87\\
NGC5614 & 10.73$\pm$0.08 & 10.67$\pm$0.04 & 15.67\\
NGC5631 & 10.20$\pm$0.08 & 10.29$\pm$0.03 & 17.44\\
NGC5633 & 9.91$\pm$0.09 & 9.93$\pm$0.04 & 12.93\\
NGC5630 & 9.37$\pm$0.09 & 9.90$\pm$0.05 & 13.78\\
NGC5657 & 9.97$\pm$0.10 & 10.13$\pm$0.04 & 11.59\\
NGC5682 & 8.87$\pm$0.09 & 9.89$\pm$0.10 & 19.63\\
NGC5720 & 10.58$\pm$0.10 & 10.66$\pm$0.02 & 11.87\\
NGC5732 & 9.77$\pm$0.10 & 10.07$\pm$0.06 & 12.28\\
UGC09476 & 9.78$\pm$0.10 & 9.91$\pm$0.03 & 15.46\\
UGC09537 & 10.57$\pm$0.09 & 11.20$\pm$0.05 & 15.76\\
UGC09542 & 9.98$\pm$0.10 & 10.31$\pm$0.03 & 12.89\\
NGC5784 & 10.75$\pm$0.08 & 10.88$\pm$0.02 & 11.91\\
NGC5797 & 10.47$\pm$0.09 & 10.49$\pm$0.02 & 13.71\\
IC1079 & 10.91$\pm$0.10 & 11.23$\pm$0.03 & 19.34\\
UGC09665 & 9.44$\pm$0.10 & 9.88$\pm$0.04 & 11.61\\
NGC5876 & 10.19$\pm$0.09 & 10.67$\pm$0.01 & 15.05\\
NGC5888 & 10.68$\pm$0.10 & 10.99$\pm$0.01 & 12.07\\
NGC5908 & 10.28$\pm$0.08 & 10.71$\pm$0.01 & 14.60\\
NGC5930 & 10.13$\pm$0.08 & 10.18$\pm$0.02 & 14.40\\
NGC5934 & 10.20$\pm$0.10 & 10.43$\pm$0.06 & 6.75\\
UGC09873 & 9.85$\pm$0.09 & 10.30$\pm$0.03 & 14.82\\
UGC09892 & 9.96$\pm$0.09 & 10.19$\pm$0.02 & 13.68\\
NGC5953 & 10.06$\pm$0.09 & 9.63$\pm$0.03 & 9.09\\
NGC5971 & 9.96$\pm$0.11 & 10.16$\pm$0.04 & 10.18\\
NGC5966 & 10.58$\pm$0.09 & 10.88$\pm$0.03 & 20.30\\
\label{tab:Gal}
\end{tabular}
\end{table}

\begin{table}[t!]
\renewcommand\thetable{A1}
      \caption{continuation}
    \label{tab:Gal}
\begin{tabular}{lccl} 
\textbf{Name} & \textbf{$M_\star$} & \textbf{$M_{dyn}$} & \textbf{$r_e$}\\
       & $[M_\odot]$ & $[M_\odot]$ & [arcsec]\\
      (1) & (2) & (3) & (4) \\
      \hline
IC4566 & 10.49$\pm$0.09 & 10.59$\pm$0.01 & 13.16\\
NGC5987 & 10.42$\pm$0.09 & 10.71$\pm$0.02 & 22.53\\
NGC5980 & 10.39$\pm$0.09 & 10.50$\pm$0.03 & 12.64\\
NGC6004 & 10.27$\pm$0.07 & 10.21$\pm$0.03 & 20.41\\
UGC10097 & 10.80$\pm$0.10 & 10.96$\pm$0.02 & 10.39\\
NGC6020 & 10.38$\pm$0.10 & 10.62$\pm$0.04 & 11.59\\
NGC6021 & 10.53$\pm$0.09 & 10.51$\pm$0.04 & 8.47\\
IC1151 & 9.49$\pm$0.09 & 9.82$\pm$0.05 & 19.34\\
UGC10123 & 9.88$\pm$0.10 & 10.29$\pm$0.02 & 11.01\\
NGC6032 & 9.83$\pm$0.10 & 10.16$\pm$0.03 & 14.79\\
NGC6060 & 10.49$\pm$0.08 & 10.66$\pm$0.01 & 20.20\\
UGC10205 & 10.69$\pm$0.11 & 10.98$\pm$0.02 & 13.95\\
NGC6063 & 9.75$\pm$0.10 & 10.05$\pm$0.03 & 17.78\\
IC1199 & 10.33$\pm$0.08 & 10.67$\pm$0.01 & 18.76\\
UGC10257 & 9.86$\pm$0.09 & 10.28$\pm$0.04 & 15.25\\
NGC6081 & 10.52$\pm$0.10 & 10.75$\pm$0.01 & 10.43\\
UGC10297 & 8.62$\pm$0.09 & 9.53$\pm$0.04 & 10.85\\
UGC10331 & 9.50$\pm$0.10 & 10.14$\pm$0.08 & 15.41\\
NGC6125 & 10.66$\pm$0.09 & 10.93$\pm$0.01 & 15.38\\
UGC10337 & 10.61$\pm$0.10 & 11.01$\pm$0.01 & 14.85\\
NGC6132 & 10.00$\pm$0.09 & 10.36$\pm$0.03 & 11.89\\
NGC6146 & 11.06$\pm$0.09 & 11.26$\pm$0.02 & 11.00\\
UGC10380 & 10.52$\pm$0.10 & 10.98$\pm$0.03 & 12.83\\
NGC6150 & 10.65$\pm$0.10 & 11.11$\pm$0.03 & 9.26\\
UGC10384 & 9.95$\pm$0.11 & 10.35$\pm$0.05 & 9.27\\
UGC10388 & 10.23$\pm$0.09 & 10.44$\pm$0.02 & 10.91\\
NGC6173 & 11.18$\pm$0.09 & 11.51$\pm$0.02 & 18.52\\
NGC6168 & 9.40$\pm$0.10 & 9.85$\pm$0.05 & 16.28\\
NGC6186 & 9.96$\pm$0.09 & 9.94$\pm$0.04 & 12.67\\
UGC10650 & 8.80$\pm$0.09 & 10.48$\pm$0.10 & 15.29\\
NGC6278 & 9.96$\pm$0.10 & 10.38$\pm$0.01 & 9.56\\
UGC10693 & 10.77$\pm$0.09 & 11.28$\pm$0.03 & 15.24\\
UGC10695 & 10.81$\pm$0.09 & 11.12$\pm$0.03 & 15.73\\
UGC10710 & 10.48$\pm$0.09 & 10.88$\pm$0.03 & 12.01\\
NGC6310 & 10.02$\pm$0.10 & 10.41$\pm$0.02 & 15.84\\
NGC6301 & 10.88$\pm$0.08 & 10.94$\pm$0.02 & 20.01\\
NGC6314 & 10.57$\pm$0.08 & 10.52$\pm$0.03 & 8.72\\
NGC6338 & 11.06$\pm$0.10 & 11.50$\pm$0.02 & 19.15\\
UGC10796 & 9.16$\pm$0.12 & 9.78$\pm$0.10 & 14.70\\
UGC10811 & 10.46$\pm$0.09 & 10.90$\pm$0.04 & 11.82\\
IC1256 & 10.19$\pm$0.09 & 10.35$\pm$0.02 & 14.60\\
NGC6394 & 10.32$\pm$0.09 & 10.68$\pm$0.02 & 9.05\\
UGC10905 & 10.81$\pm$0.09 & 11.04$\pm$0.03 & 12.37\\
NGC6411 & 10.54$\pm$0.09 & 10.77$\pm$0.03 & 17.82\\
NGC6427 & 10.00$\pm$0.10 & 10.30$\pm$0.02 & 8.88\\
UGC10972 & 10.22$\pm$0.10 & 10.51$\pm$0.03 & 19.31\\
NGC6478 & 10.83$\pm$0.09 & 10.97$\pm$0.02 & 17.36\\
NGC6497 & 10.52$\pm$0.08 & 10.74$\pm$0.01 & 11.99\\
NGC6515 & 10.79$\pm$0.09 & 10.97$\pm$0.04 & 13.07\\
UGC11228 & 10.55$\pm$0.10 & 10.48$\pm$0.03 & 7.23\\
NGC6762 & 9.91$\pm$0.09 & 10.03$\pm$0.06 & 11.21\\
MCG-02-51-004 & 10.66$\pm$0.09 & 10.75$\pm$0.01 & 15.80\\
NGC6941 & 10.64$\pm$0.09 & 10.72$\pm$0.01 & 15.04\\
NGC6945 & 9.11$\pm$0.08 & 10.41$\pm$0.02 & 12.80\\
NGC6978 & 10.49$\pm$0.10 & 10.73$\pm$0.01 & 12.15\\
UGC11649 & 10.14$\pm$0.08 & 10.28$\pm$0.02 & 14.54\\
UGC11680NED01 & 10.84$\pm$0.09 & 10.94$\pm$0.01 & 14.56\\
\label{tab:Gal}
\end{tabular}
\end{table}

\begin{table}[t!]
\renewcommand\thetable{A1}
      \caption{continuation}
    \label{tab:Gal}
\begin{tabular}{lccl} 
\textbf{Name} & \textbf{$M_\star$} & \textbf{$M_{dyn}$} & \textbf{$r_e$}\\
       & $[M_\odot]$ & $[M_\odot]$ & [arcsec]\\
      (1) & (2) & (3) & (4) \\
      \hline
NGC7025 & 10.82$\pm$0.09 & 11.05$\pm$0.02 & 18.20\\
NGC7047 & 10.51$\pm$0.10 & 10.70$\pm$0.02 & 19.97\\
UGC11717 & 10.66$\pm$0.09 & 10.44$\pm$0.17 & 11.77\\
MCG-01-54-016 & 8.38$\pm$0.11 & 9.82$\pm$0.07 & 12.07\\
NGC7194 & 10.75$\pm$0.09 & 11.20$\pm$0.05 & 11.96\\
UGC12054 & 8.58$\pm$0.10 & 9.44$\pm$0.07 & 10.41\\
NGC7311 & 10.59$\pm$0.09 & 10.59$\pm$0.02 & 10.60\\
NGC7321 & 10.65$\pm$0.08 & 10.74$\pm$0.03 & 12.04\\
UGC12127 & 11.01$\pm$0.09 & 11.46$\pm$0.04 & 18.86\\
NGC7364 & 10.60$\pm$0.10 & 10.57$\pm$0.02 & 10.56\\
UGC12185 & 10.36$\pm$0.08 & 10.57$\pm$0.04 & 9.71\\
NGC7436B & 11.21$\pm$0.10 & 11.52$\pm$0.02 & 22.59\\
UGC12274 & 10.64$\pm$0.09 & 10.85$\pm$0.02 & 12.39\\
UGC12308 & 8.95$\pm$0.10 & 10.03$\pm$0.07 & 20.19\\
NGC7466 & 10.55$\pm$0.09 & 10.63$\pm$0.02 & 12.64\\
NGC7489 & 10.53$\pm$0.08 & 10.42$\pm$0.04 & 16.66\\
NGC7550 & 10.87$\pm$0.09 & 11.07$\pm$0.03 & 16.66\\
NGC7549 & 10.38$\pm$0.09 & 10.60$\pm$0.02 & 16.64\\
NGC7563 & 10.28$\pm$0.09 & 10.59$\pm$0.01 & 8.88\\
NGC7562 & 10.68$\pm$0.09 & 10.85$\pm$0.03 & 14.03\\
NGC7591 & 10.62$\pm$0.10 & 10.51$\pm$0.03 & 13.08\\
UGC12494 & 9.33$\pm$0.10 & 10.09$\pm$0.17 & 14.49\\
IC5309 & 10.47$\pm$0.11 & 10.05$\pm$0.07 & 13.31\\
NGC7608 & 9.13$\pm$0.09 & 9.72$\pm$0.03 & 7.61\\
NGC7611 & 10.20$\pm$0.11 & 10.71$\pm$0.07 & 9.76\\
UGC12519 & 9.84$\pm$0.09 & 10.12$\pm$0.03 & 11.73\\
NGC7619 & 10.80$\pm$0.08 & 11.15$\pm$0.02 & 21.44\\
NGC7623 & 10.20$\pm$0.09 & 10.14$\pm$0.03 & 7.97\\
NGC7625 & 9.64$\pm$0.08 & 9.51$\pm$0.03 & 9.80\\
NGC7631 & 10.21$\pm$0.09 & 10.28$\pm$0.02 & 14.10\\
NGC7653 & 10.46$\pm$0.09 & 10.17$\pm$0.04 & 12.28\\
NGC7671 & 10.31$\pm$0.10 & 10.60$\pm$0.02 & 9.26\\
NGC7683 & 10.46$\pm$0.10 & 10.69$\pm$0.03 & 12.80\\
NGC7684 & 10.47$\pm$0.10 & 10.48$\pm$0.02 & 9.04\\
NGC7691 & 10.25$\pm$0.09 & 10.21$\pm$0.03 & 22.82\\
NGC7711 & 10.53$\pm$0.09 & 10.69$\pm$0.03 & 13.47\\
NGC7716 & 10.17$\pm$0.08 & 9.94$\pm$0.04 & 14.16\\
NGC7722 & 10.74$\pm$0.09 & 10.91$\pm$0.07 & 18.01\\
UGC12723 & 9.78$\pm$0.13 & 10.27$\pm$0.02 & 15.94\\
NGC7738 & 10.67$\pm$0.10 & 10.64$\pm$0.01 & 11.51\\
UGC12810 & 10.43$\pm$0.09 & 10.72$\pm$0.02 & 13.56\\
UGC12816 & 9.74$\pm$0.11 & 10.29$\pm$0.05 & 12.85\\
NGC7783NED01 & 10.64$\pm$0.10 & 11.08$\pm$0.03 & 9.98\\
NGC7787 & 10.56$\pm$0.10 & 10.59$\pm$0.04 & 13.55\\
UGC12857 & 9.55$\pm$0.10 & 9.83$\pm$0.06 & 18.39\\
UGC12864 & 9.69$\pm$0.10 & 9.95$\pm$0.06 & 14.00\\
NGC7800 & 8.93$\pm$0.08 & 10.10$\pm$0.07 & 16.30\\
NGC5947 & 10.28$\pm$0.10 & 10.26$\pm$0.03 & 10.55\\
NGC5947 & 10.28$\pm$0.10 & 10.26$\pm$0.03 & 10.55\\
NGC5947 & 10.28$\pm$0.10 & 10.26$\pm$0.03 & 10.55\\
NGC0180 & 10.72$\pm$0.08 & 10.59$\pm$0.04 & 20.19\\
NGC0192 & 10.46$\pm$0.09 & 10.49$\pm$0.02 & 13.90\\
NGC0155 & 10.72$\pm$0.10 & 10.82$\pm$0.02 & 13.55\\
NGC0160 & 10.73$\pm$0.09 & 10.87$\pm$0.01 & 19.42\\
NGC0169 & 10.88$\pm$0.10 & 10.98$\pm$0.06 & 19.42\\
NGC0171 & 10.29$\pm$0.08 & 9.89$\pm$0.04 & 15.81\\
NGC0177 & 10.35$\pm$0.10 & 10.45$\pm$0.04 & 17.53\\
\label{tab:Gal}
\end{tabular}
\end{table}
\clearpage

\section{Linear fit parameters and scatters
for the $M_\star-S_{0.5}$ correlations.}
\label{Appendix: AppendixB}

All scatters are estimated from the linear fit as the standard deviation of all residuals, we consider stellar mass, $M_\star$, as independent variable. $log(S_{0.5})\ =\ a\ +\ blog(M_\star)$. $S_{0.5}$ is given in [km/s] and \ $M_\star$ in $M_\odot$.

\begin{table}[h]
\begin{tabular}{llcl}
\hline
\hline
Fit & zero-point(a) & slope(b) & scatter \\
\hline
\textbf{Integrated kinematics} & & & \\
\textit{Gas} & & & \\
Forward & $-0.95\pm0.11$ & $0.29\pm0.01$ & 0.087\\
Inverse & $-1.99\pm0.18$ & $0.38\pm0.01$ & 0.100\\
Bisector & $-1.46\pm0.13$ & $0.33\pm0.01$ & 0.090\\
Orthogonal & $-1.03\pm0.12$ & $0.29\pm0.01$ & 0.087\\
\textit{Stellar} & & & \\
Forward & $-0.61\pm0.10$ & $0.25\pm0.01$ & 0.075\\
Inverse & $-1.52\pm0.14$ & $0.34\pm0.01$ & 0.086\\
Bisector & $-1.06\pm0.11$ & $0.29\pm0.01$ & 0.078\\
Orthogonal & $-0.67\pm0.10$ & $0.26\pm0.01$ & 0.075\\
\textit{Total (Gas+Stellar)} & & & \\
Forward & $-0.72\pm0.07$ & $0.26\pm0.01$ & 0.082\\
Inverse & $-1.73\pm0.11$ & $0.36\pm0.01$ & 0.096\\
Bisector & $-1.22\pm0.08$ & $0.31\pm0.01$ & 0.086\\
Orthogonal & $-0.79\pm0.07$ & $0.27\pm0.01$ & 0.082\\
\hline
\textbf{Resolved kinematics} & & & \\
\textit{Gas} & & & \\
Forward & $-0.89\pm0.14$ & $0.28\pm0.01$ & 0.053\\
Inverse & $-1.29\pm0.18$ & $0.32\pm0.02$ & 0.056\\
Bisector & $-1.09\pm0.15$ & $0.30\pm0.01$ & 0.054\\
Orthogonal & $-0.92\pm0.13$ & $0.29\pm0.01$ & 0.053\\
\textit{Stellar} & & & \\
Forward & $-0.67\pm0.12$ & $0.26\pm0.01$ & 0.052\\
Inverse & $-1.40\pm0.17$ & $0.33\pm0.01$ & 0.056\\
Bisector & $-1.03\pm0.14$ & $0.29\pm0.01$ & 0.050\\
Orthogonal & $-0.72\pm0.12$ & $0.27\pm0.01$ & 0.052\\
\textit{Total (gas+stellar)} & & & \\
Forward & $-0.66\pm0.08$ & $0.26\pm0.01$ & 0.054\\
Inverse & $-1.31\pm0.13$& $0.32\pm0.01$ & 0.059\\
Bisector & $-0.98\pm0.10$& $0.29\pm0.01$ & 0.055\\
Orthogonal & $-0.71\pm0.11$ & $0.27\pm0.01$ & 0.054\\
\hline
\textbf{Only spiral galaxies} & & & \\
\textit{Gas} & & & \\
Forward & $-0.87\pm0.15$ & $0.28\pm0.02$ & 0.043 \\
Inverse & $-0.98\pm0.11$ & $0.29\pm0.01$ & 0.044 \\
Bisector & $-0.93\pm0.14$ & $0.29\pm0.01$ & 0.043 \\
Orthogonal & $-0.88\pm0.09$ & $0.28\pm0.01$ & 0.043 \\
\textit{Stellar} & & & \\
Forward & $-0.88\pm0.10$ & $0.28\pm0.01$ & 0.051 \\
Inverse & $-1.35\pm0.12$ & $0.32\pm0.02$ & 0.057 \\
Bisector & $-1.12\pm0.13$ & $0.30\pm0.01$ & 0.054 \\
Orthogonal & $-0.92\pm0.12$ & $0.28\pm0.01$ & 0.052\\
\textit{Total (gas+stellar)} & & & \\
Forward & $-0.77\pm0.12$ & $0.27\pm0.01$ & 0.051 \\
Inverse & $-1.16\pm0.10$ & $0.31\pm0.01$ & 0.056 \\
Bisector & $-0.96\pm0.16$ & $0.29\pm0.01$ & 0.053 \\
Orthogonal & $-0.80\pm0.11$ & $0.27\pm0.01$ & 0.051\\
\end{tabular}
\end{table}

\bsp	
\label{lastpage}
\end{document}